\documentclass[journal]{IEEEtran}

\ifCLASSINFOpdf
\else
\fi

\usepackage{cite}
\usepackage{indentfirst}
\usepackage{graphicx}
\usepackage{amssymb}
\usepackage{amsfonts}
\usepackage{amsmath}
\usepackage{cases}
\usepackage{algpseudocode}
\usepackage{subfigure}
\usepackage{mathrsfs}
\usepackage{makecell}
\usepackage[linesnumbered,boxed,ruled,commentsnumbered]{algorithm2e}
\SetKwRepeat{Do}{do}{while}%
\usepackage[ruled]{algorithm2e}
\usepackage{amsthm}
\usepackage{multirow}
\usepackage{url}
\usepackage{amssymb}
\usepackage{color}
\usepackage{textcomp}
\usepackage{afterpage}

\newtheoremstyle{baitingstyle}{3pt}{3pt}{\itshape}{1.2em}{\itshape}{:}{0.5em}{{\thmname{#1}\thmnumber{ #2}\thmnote{ (#3)}}}
\theoremstyle{baitingstyle}
\newtheorem{theorem}{Theorem}
\theoremstyle{plain}

\theoremstyle{baitingstyle}

\theoremstyle{baitingstyle}
\newtheorem{remark}{Remark}
\theoremstyle{baitingstyle}

\theoremstyle{baitingstyle}

\theoremstyle{baitingstyle}
\newtheorem{proposition}{Proposition}
\newcommand*{\QEDA}{\hfill\ensuremath{\blacksquare}} 
\newenvironment{proofproposition1}
{{\indent\itshape Proof of Proposition 1:}\,}{\hfill$\QEDA$}

\newenvironment{prooftheorem1}
{{\indent\itshape Proof of Theorem 1:}\,}{\hfill$\QEDA$}

\renewenvironment{proof}{{\indent\indent\itshape Proof:}\,}{\hfill$\QEDA$}

\begin{document}
%
\title{Distributed Charging Coordination for Electric Trucks under Limited Facilities and \\Travel Uncertainties}
\author{Ting~Bai$^{1}$,~Yuchao~Li$^{2}$,~Andreas A.~Malikopoulos$^{3}$,
        Karl~H.~Johansson$^{1}$,~\IEEEmembership{Fellow,~IEEE,}
        and~Jonas~M{\aa}rtensson$^{1}$
\thanks{This work was supported in part by the Swedish Research Council Distinguished Professor (Grant Number: 2017-01078), the Knut and Alice Wallenberg Foundation, the Swedish Strategic Research Foundation CLAS (Grant Number: RIT17-0046), NSF (Grant Numbers: CNS-2401007, CMMI-2219761, IIS-2415478), and MathWorks.}\vspace{1.5pt}

\thanks{$^{1}$T. Bai, K. H. Johansson, and J. Mårtensson are with the Integrated Transport Research Lab, the Division of Decision and Control Systems, and the Digital Futures, KTH Royal Institute of Technology, SE-10044 Stockholm, Sweden (e-mails: tingbai@kth.se; kallej@kth.se; jonas1@kth.se).}
\thanks{$^{2}$Y. Li is with the School of Computing and Augmented Intelligence, Arizona State University, AZ-85281 Tempe, the United States (e-mail: {yuchaoli@asu.edu}).}
\thanks{$^{3}$A. A. Malikopoulos is with the School of Civil and Environmental Engineering, Cornell University, NY-14853 Ithaca, the United States (e-mail: {amaliko@cornell.edu}).}
}

\markboth{}%
{Shell \MakeLowercase{\textit{et al.}}: Bare Demo of IEEEtran.cls for IEEE Journals}
\maketitle

\begin{abstract}
In this work, we address the problem of charging coordination between electric trucks and charging stations. The problem arises from the tension between the trucks' nontrivial charging times and the stations' limited charging facilities. Our goal is to reduce the trucks' waiting times at the stations while minimizing individual trucks' operational costs. We propose a distributed coordination framework that relies on computation and communication between the stations and the trucks, and handles uncertainties in travel times and energy consumption. Within the framework, the stations assign a limited number of charging ports to trucks according to the first-come, first-served rule. In addition, each station constructs a waiting time forecast model based on its historical data and provides its estimated waiting times to trucks upon request. When approaching a station, a truck sends its arrival time and estimated arrival-time windows to the nearby station and the distant stations, respectively. The truck then receives the estimated waiting times from these stations in response, and updates its charging plan accordingly while accounting for travel uncertainties. We performed simulation studies for $1,000$ trucks traversing the Swedish road network for $40$ days, using realistic traffic data with travel uncertainties. The results show that our method reduces the average waiting time of the trucks by $46.1\%$ compared to offline charging plans computed by the trucks without coordination and update, and by $33.8\%$ compared to the coordination scheme assuming zero waiting times at distant stations. 
\end{abstract}

\vspace{-30pt}

\IEEEpeerreviewmaketitle

\section*{Nomenclature}\label{Section 0}
\vspace{-20pt}
\subsection{Sets}
\vspace{-10pt}
\begin{tabular}{ll}
  $\Re_{+}$ & Set of nonnegative reals.\vspace{1.5pt}\\
  $C$ & Set of charging ports.
\end{tabular}

\subsection{Parameters}
\vspace{4pt}
\begin{tabular}{ll}
  $N$ & Number of charging stations.\vspace{1.5pt}\\
  $d_{\ell}$ & Detour time between ramp $r_{\ell}$ and station $S_{\ell}$.\vspace{1.5pt}\\
  $\tau_{\ell}$ & Nominal travel time from ramp $r_{\ell}$ to ramp $r_{\ell+1}$.\vspace{1.5pt}\\
  $\delta_{\tau,\ell}$ & Random fluctuation in travel time $\tau_{\ell}$.\vspace{1.5pt}\\
  $\delta_{e,\ell}$ & Random fluctuation in battery consumption \vspace{1.5pt}\\
  & from ramp $r_{\ell}$ to ramp $r_{\ell+1}$.\vspace{1.5pt}\\
  $\delta_{\ell}^1$/$\delta_{\ell}^2$ & Nonnegative constants used to characterize travel\vspace{1.5pt} \\
  &time/energy consumption uncertainty.\vspace{1.5pt}\\
  $e_{\text{ini}}$ & Initial battery of a truck.\vspace{1.5pt}\\
  $\bar{P}$ & Battery consumption per travel time unit.\vspace{1.5pt}\\
  $P_{\ell}$ & Charging power provided by station $S_{\ell}$.\vspace{1.5pt}\\
  $P_{\max}$ & Maximum charging power acceptable to a truck.\vspace{1.5pt}\\
  $e_s$ & Safe margin on a truck's remaining battery.\vspace{1.5pt}\\
  $e_f$ & Full battery of a truck. \vspace{1.5pt}\\
  $a_{\text{ini}}$ & Departure time of a truck from its origin.\vspace{1.5pt}\\
  $a_{\text{end}}$ & Latest arrival time indicated by the deadline.\vspace{1.5pt}\\
  $\xi$ & Labor cost per time unit.\vspace{1.5pt}\\
  $\theta_{\ell}$ & Electricity cost per charging time at station $S_{\ell}$.\vspace{1.5pt}\\
  $\gamma$ & Tuning penalty parameter for deadline violation.
\end{tabular}
\vspace{-12pt}
\subsection{Variables}
\vspace{4pt}
\begin{tabular}{ll}
  $b_{\ell}$ & Binary charging decision at station $S_{\ell}$.\vspace{1.5pt}\\
  $t_{\ell}$ & Charging time at station $S_{\ell}$.\vspace{1.5pt}\\
  $\tilde{w}_{\ell}$ & Estimated waiting time provided by station $S_{\ell}$.\vspace{1.5pt}\\
  $\bar{w}_{\ell}$ & Estimated maximum waiting time at station $S_{\ell}$.\vspace{1.5pt}\\
  $t_{a,c}$ & Earliest time since when port $c$ is available.\vspace{1.5pt}\\
  $a_{\ell}$ & A truck's arrival time at ramp $r_{\ell}$.\vspace{1.5pt}\\
  $e_{\ell}$ & A truck's remaining battery when first arriving \vspace{1.5pt}\\
  & at ramp $r_{\ell}$.\vspace{1.5pt}\\
  $\Delta{e}_{\ell}$ & Battery charged at station $S_{\ell}$.\vspace{1.5pt}\\
  $\Delta{\bar{e}}_{\ell}$ & Maximum increased battery at station $S_{\ell}$.\vspace{1.5pt}\\
  $w_{\ell}$ & A truck's waiting time at station $S_{\ell}$.\vspace{1.5pt}\\
  $\underline{a}_{\ell}$/$\bar{a}_{\ell}$ & Earliest/latest arrival time of a truck at ramp $r_{\ell}$.\vspace{1.5pt}\\
  $\underline{a}_{\ell}^k$/$\bar{a}_{\ell}^k$ & Earliest/latest arrival time at ramp $r_{\ell}$ computed\vspace{1.5pt}\\
  & at ramp $r_k$.
  \end{tabular}

\section{Introduction}\label{Section I}
\IEEEPARstart{T}{he} move towards vehicle electrification has gained global momentum as a solution to climate change and energy shortages~\cite{bao2021global}. Given that road freight transportation contributes significantly to emissions both in Europe~\cite{siskos2019assessing,10209062} and globally~\cite{ritchie2023sector}, replacing diesel trucks with electric ones can yield considerable benefits. To facilitate this transition, rapid progress has been made in many related aspects, including battery technologies \cite{burd2021improvements}, infrastructure development~\cite{metais2022too,gupta2021optimal}, and incentive policy designs~\cite{yan2018economic,breetz2018electric}. Despite these efforts, several obstacles still impede the widespread adoption of electric trucks.

One such issue is known as \emph{range anxiety}~\cite{yong2023electric}, which describes drivers' concerns about insufficient battery power to reach their destinations. This problem is especially relevant for trucks on long-range delivery missions, where even fully charged batteries may not cover the entire distance~\cite{wassiliadis2021review}. Consequently, it is often necessary to plan where and for how long to recharge the truck at available charging stations. To address such a problem and the associated range anxiety, there has been a growing body of research in developing charging planning methods for trucks. As a part of problem simplification, much of the work assumes sufficient charging facilities at the given charging stations. For instance, \cite{zahringer2022time} and \cite{10147895} designed charging planning methods for individual trucks that assume no additional waiting time due to charging congestion. These methods also take into consideration mandatory rest regulations for the drivers. Other studies, such as \cite{schoenberg2022reducing,erdelic2019survey,sweda2012finding}, formulated the charging planning problems as the shortest path problems, where the minimizing objectives vary from the consumed energy, the total travel time, to the operational cost. The computational study given in \cite{kin2021different} examined various charging strategies for large vehicle fleets, where the target is to minimize the total cost of the fleets. 

Despite the success reported in these works, the assumption of sufficient charging facilities is often overly optimistic. In practice, charging stations have a limited number of charging ports, and the charging times for trucks can be long. When stations are used by trucks from different carriers and there is no coordination, trucks may get stuck in prolonged waiting queues, resulting in increased travel time and labor costs, and potential violations of delivery deadlines. To facilitate the wider adoption of electric trucks despite limited charging resources, there has been much research on coordinated charging strategies in recent years~\cite{8855113,hu2013coordinated,mediwaththe2018game,tang2014online,kisacikoglu2017distributed,karfopoulos2012multi}. Depending on the party responsible for coordination, these methods can be categorized into three types: station-based methods, truck-based methods, and holistic methods.

In station-based methods, charging stations direct or suggest appropriate sites for electric trucks to charge. The coordination objectives of these methods vary from maximizing the utilization of charging infrastructure~\cite{6486021}, mitigating power overloads across stations~\cite{elghitani2020efficient}, enhancing the operational flexibility of power systems~\cite{zhang2021distributed}, to minimizing operational costs at charging stations~\cite{gupta2020collaborative}, to name only a few. In contrast to approaches that focus on a single optimization objective, some works~\cite{qian2021multi,lee2018analysis} model the charging coordination problem as non-cooperative games. This formulation allows individual stations to maximize their own revenues by offering competitive charging prices to vehicles. Regardless of the number of objectives optimized, these station-based methods overlook the costs and payoffs of the trucks. 

In contrast to the station-based methods, truck-based methods prioritize the interests of the trucks. In these schemes, trucks compute their charging plans with little support from the stations. For instance, the method developed in \cite{del2016smart} aims to minimize the total travel time of individual vehicles. To account for limited charging facilities, this method employs data-based queuing models to represent other trucks competing for the charging ports. As a result, the truck need not communicate with the stations when scheduling charging plans. Similar to this work, the methods developed in~\cite{qin2011charging,yang2013charge} focus on minimizing vehicles' extra waiting times at stations, where they use probabilistic models to describe the arrival process of competing trucks. Recent work~\cite{qian2025empirical} constructed three charging strategies for electric trucks: spontaneous, prospect theory-based, and reinforcement learning-based charging methods, where the charging queues are modeled using queuing theory algorithms. Despite the simplicity of these schemes, the lack of communication between the trucks and the stations may lead to potential performance deterioration and travel-related uncertainties are not considered.

As an alternative to the aforementioned two kinds of schemes, the holistic methods coordinate the charging behaviors of electric trucks via centralized computation, using information received from both the trucks and the stations. These approaches are suitable for scenarios where trucks and charging stations belong to the same interested party, such as big fleet owners. As a result, a common objective shared by all parties involved is optimized through coordination. For instance, \cite{al2022smart} proposed a smart return-to-base charging strategy for electric trucks, where chargers are installed at only the starting and ending points of the trips. The problem was modeled as a nonlinear optimization aimed at minimizing peak energy demand at the facility while ensuring compliance with the operational schedules of the truck fleet. Methods belonging to this type can also be found in \cite{tang2020congestion,8884678}. Apart from the restrictions on the applicable scenarios, these centralized methods typically require extensive communication and full control of each component involved in the system. 

In this work, we address the coordination problem where each truck operates based on its own interests. Moreover, we take into consideration the uncertainties in travel times and energy consumption, which are prevalent in practice. Consequently, both station-based methods and holistic coordination schemes cannot be directly applied. While truck-based methods may be used for such problems, our proposed framework enhances coordination performance by incorporating communication between individual trucks and stations, eliminating the need for probabilistic assumptions about truck arrival patterns at the stations. Building upon our previous work~\cite{bai2024distributed}, the proposed framework involves limited communication between trucks and the stations along their routes, and features charging planning methods that handle bounded travel uncertainties. The main contributions of this work are summarized as follows.
\begin{itemize}
    \item We propose a distributed charging coordination framework that reduces trucks' waiting times at stations while minimizing their operational costs. The framework involves limited communication between charging stations and trucks, enabling individual trucks to optimize their charging plans dynamically and independently.
    \vspace{1.5pt}
    
    \item We design a communication and computation scheme for the stations and trucks that facilitates the estimation of trucks' waiting times at stations. The scheme involves waiting time forecast models constructed by stations and estimated arrival-time windows computed by trucks. It decouples waiting time estimation from charging plan computation, leading to effective charging planning.
    \vspace{1.5pt}
    
    \item We propose a distributed charging planning approach for individual trucks, which handles bounded uncertainties in travel times and energy consumption while ensuring the feasibility of the obtained charging plans.
    \vspace{1.5pt}

    \item We demonstrate through analysis that the estimated arrival-time window of a truck at each distant station shrinks as the truck approaches the station. Moreover, we provide an upper bound on the additional cost due to waiting time estimate errors, which decreases as the truck travels toward its destination.
\end{itemize}

Compared to our previous work \cite{bai2024distributed}, the present framework introduces three new components: 1) waiting time forecast models computed by the stations, 2) estimated arrival-time windows computed by the trucks, and 3) a new charging planning method for the trucks with feasibility guarantees despite travel uncertainties. To test the effectiveness of the proposed coordination framework, we conducted simulation studies for $1,000$ trucks traversing the Swedish road network over $40$ days with bounded travel uncertainties. In such cases, the coordination scheme in \cite{bai2024distributed} cannot guarantee the feasibility of charging plans. In contrast, our new coordination scheme consistently provides feasible solutions and achieves about $33.8\%$ reduction in waiting time compared to the coordination scheme~\cite{bai2024distributed}, due to the integration of component 3). Furthermore, it reduces the average waiting time for each truck by approximately $46.1\%$ compared to offline charging plans computed without coordination. These results demonstrate that the new components developed in this work contribute to improved coordination performance and the capability to handle travel uncertainties.

The remainder of the paper is structured as follows. Section~\ref{Section II} provides an overview of the coordination framework. Section~\ref{Section III} delves into the scheduling mechanisms carried out by charging stations, including charging resource allocation and waiting time computation. Section~\ref{Section IV} introduces the approach to computing optimal charging plans for individual trucks, while Section~\ref{Section V} discusses the simulation studies using realistic road and traffic data. Finally, Section~\ref{Section VI} presents concluding remarks and possible directions for future research. 

\section{Problem Description and Overview of the Coordination Framework} \label{Section II}
This section provides a description of the charging coordination problem to be addressed in this work and an overview of the proposed coordination framework.

\subsection{Problem Description}
We consider a large collection of electric trucks traversing a road network, where trucks may need to charge midway due to limited battery capacities and long travel distances. In many cases, the charging times required by the trucks can be long, and the charging resources at stations are limited. Therefore, without coordination, trucks may find themselves stuck in long waiting queues for charging ports. To reduce waiting times and mitigate charging congestion at stations, it is essential to establish effective coordination schemes for both the stations and trucks. In particular, throughout this paper, we consider the coordination problem with the following settings:
\begin{itemize}
    \item[a)] There are a fixed number of charging stations in the road network, each with a fixed number of charging ports. 
    \item[b)] For all the trucks traversing the road network, their routes are pre-planned. In addition, heading to charging stations may lead to detours.
    \item[c)] The traveling time and energy consumption over the pre-planned routes are subject to bounded uncertainties, whereas those for the detours are small and assumed to be deterministic.
    \item[d)] Each truck operates for its own interests.
\end{itemize}

Given the conditions stated above, the charging coordination problem consists of designing communication schemes between the trucks and stations, developing appropriate charging port assignment mechanisms at the stations, and proposing effective charging planning methods for the trucks. The goal is to minimize the labor and charging costs for individual trucks while reducing their waiting times at the stations. 

\subsection{Overview of the Coordination Framework}
The proposed coordination framework involves a little communication between trucks and charging stations along their routes. Within the framework, each station assigns limited charging facilities to trucks upon their arrival and provides estimated waiting times to the trucks in response to their requests. Based on the information received from the charging stations, trucks optimize their charging plans independently while accounting for bounded travel uncertainties. The computations performed by stations and trucks, as well as the information exchanges between them, are described below.
\begin{figure}[t!]
     \centering
     \includegraphics[width=0.9\linewidth]{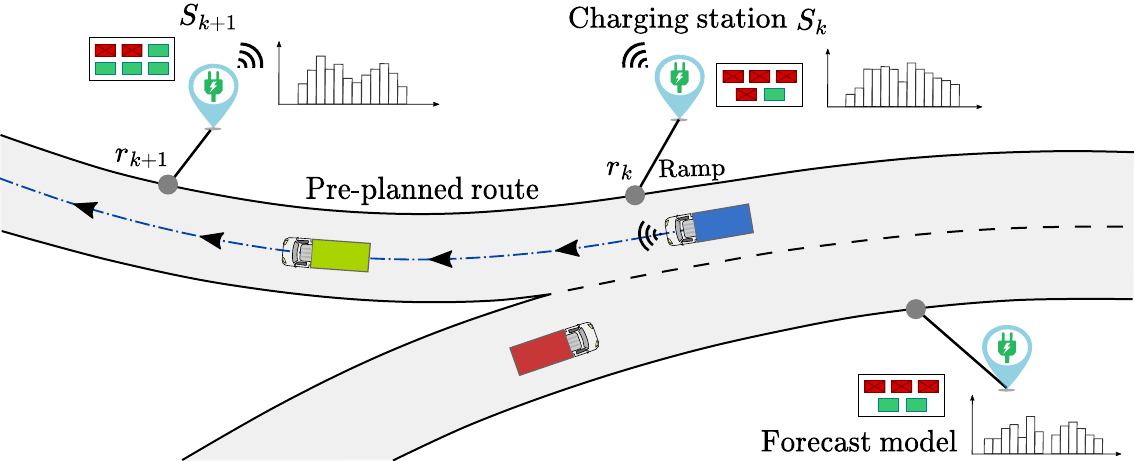}
     \vspace{-2pt}
     \caption{The road network and communication scheme between trucks and charging stations. Taking the blue truck as an example, it communicates with the charging stations along its route each time it reaches a ramp. Charging stations have limited facilities, where occupied charging ports are indicated by red blocks and available ones by green. Additionally, each station maintains a forecast model to estimate waiting times for trucks.}
      \label{Fig.1}
   \end{figure}

\subsubsection{Computation by the Stations} As illustrated in Figure~\ref{Fig.1}, the main roads of the traffic network are divided into road segments by \emph{ramps} (denoted by $r_k$ and $r_{k+1}$ in Figure~\ref{Fig.1}, with the connected charging stations denoted by $S_k$ and $S_{k+1}$, respectively), which serve as access points to individual charging stations with the shortest detours. Within our framework, stations assign their charging ports to the arrived trucks following the \textit{first-come, first-served rule}. In addition, stations provide waiting time estimations to trucks upon request to facilitate the charging plan computations. 

In particular, each station computes the charging ports assigned to arrived trucks, and waiting time estimations upon request by trucks. While computing charging port indices is straightforward, estimating waiting times for distant trucks (i.e., trucks that have not yet reached the ramp connected to the station) is difficult. The main challenges arise from:
\begin{itemize}
\item[C1)] The charging schedule at each station can be dynamically changed with the arrival and departure of individual trucks. Without the knowledge of other trucks' charging plans, it is difficult for the stations to accurately estimate waiting times for distant trucks, even given the trucks' precise arrival time information.
\item [C2)] The arrival time of a distant truck at the station depends on the truck's charging decisions and the actual waiting times at all preceding stations. This, in turn, affects the estimated waiting times received by the truck and its charging decisions at preceding stations. Such couplings make it hard for the truck to provide a precise arrival time when requesting waiting time estimations.
\end{itemize}
  
To address challenge C1), each station operates in two phases in our framework: the data collection phase and the nominal operation phase. In the data collection phase, the station records the actual arrival and waiting times of trucks locally and constructs a waiting time forecast model based on these data. During this period, the waiting time estimates $\tilde{w}$ sent to distant trucks are computed based on the available data.\footnote{During the initial data collection phase, in the absence of historical data and prior knowledge about charging congestion patterns, a simplified assumption is adopted: the waiting time estimate is set to zero for all distant stations.} The model takes the arrival time as input and outputs the estimated waiting time. Once the forecast model is obtained, it is used to compute waiting time estimates without the information of other trucks. 

The challenge C2) arises from the coupling between the waiting time estimates and the charging plan. This is resolved through the interaction between the waiting time forecast models and the trucks, as we will discuss next.

\subsubsection{Computation by the Trucks} For a given truck, let $N$ represent the number of charging stations along its route. Consequently, there are $N$ ramps in the route, each connecting to a distinct station. The truck computes its charging plan each time it reaches a ramp, denoted as $r_k$, $k\!=\!1,\dots,N$. The target is to minimize its own operational cost to traverse the remaining route for delivery mission completion. Upon reaching ramp $r_k$, the charging decisions of the truck are described by the collection of variables $b_{\ell}\!\in\!\{0,1\}$, $t_{\ell}\!\in\!{\Re_+}$, $\ell\!=\!k,\dots,N$, where $b_{\ell}$ represents whether to charge the truck at its $\ell$th station, and $t_{\ell}$ represents its planned charging time if $b_{\ell}\!=\!1$, with $\Re_+$ denoting the set of nonnegative reals. The local information available for the truck to compute these values includes the travel times on its main route and for detours, the charging and battery consumption models, the bounds on travel uncertainties, and other related information to be detailed in Section~\ref{Section IV}. The critical information provided by the stations $S_{\ell}$ is the truck's waiting time estimation, denoted by $\Tilde{w}_{\ell}\!\in\!\Re_+,\, \ell\!=\!k\!+\!1,\dots,N$. 

As discussed earlier, the coupling between the waiting time estimates and the charging plan presents a major challenge; see C2) above. While the waiting time forecast models at stations can be helpful, they rely on the estimated arrival times provided by the trucks. In our framework, each truck facilitates the waiting time estimates at each station along its route by computing an \emph{estimated arrival-time window}. In particular, upon reaching ramp $r_k$, the truck computes the estimated arrival-time windows for each of its distant stations $S_\ell$, $\ell\!=\!k\!+\!1,\dots,N$ (excluding the nearby station $S_k$). The computation assumes the best and worst scenarios for related variables, thereby decoupling the charging planning from the arrival time estimations.
\begin{figure}[t]
     \centering
     \includegraphics[width=0.9\linewidth]{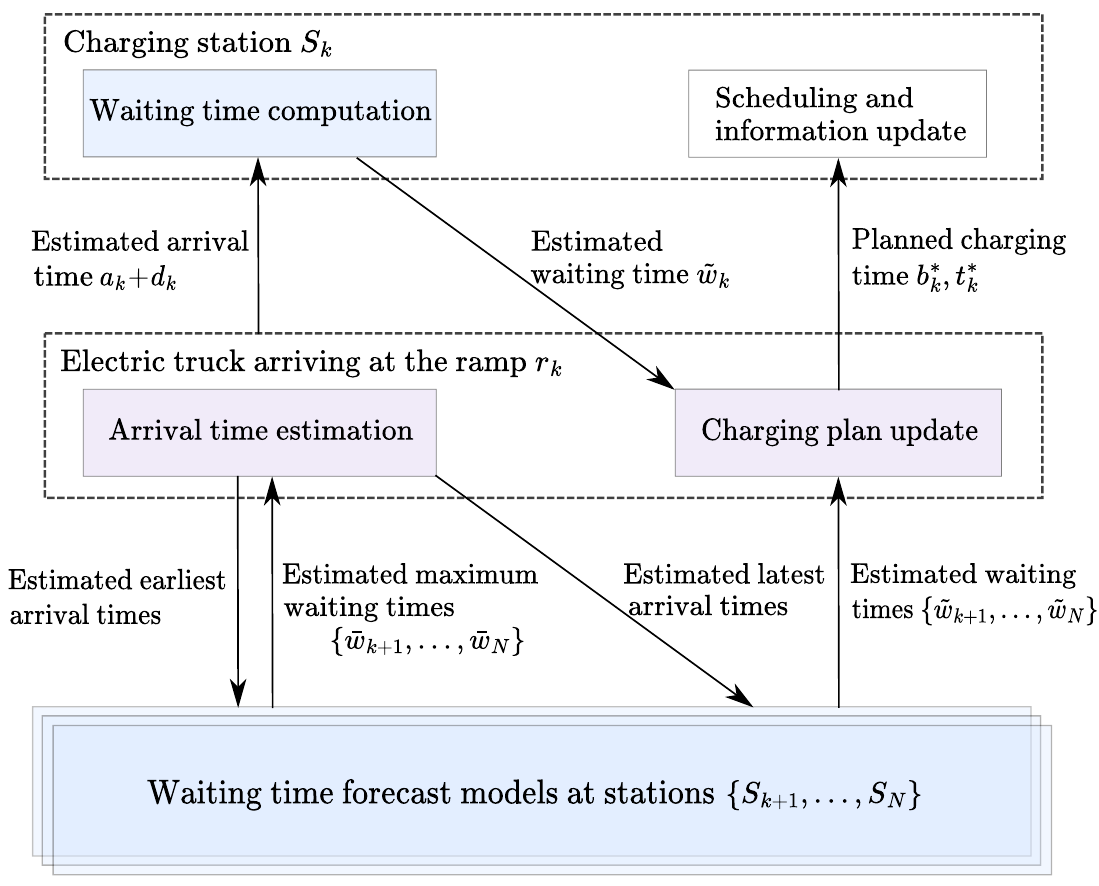}
     \vspace{-8pt}
      \caption{An overview of the communication and coordination framework between each truck and the charging stations.}
      \label{Fig.2}
   \end{figure}
\subsubsection{Communication and Coordination Architecture} We are now ready to present an overview of the communication and coordination architecture. Suppose that all the stations have computed their waiting time forecast models. For each truck traversing the road network, the following sequence of communications and computations occur whenever the truck reaches the $k$th ramp, $k\!=\!1,\dots,N$, in its pre-planned route.   
\begin{itemize}
    \item[1)] The truck sends to station $S_k$ connected to $r_k$ its estimated arrival time, which is the present time $a_k$ plus the detour time $d_k$ needed for reaching the station.
    \item[2)] The station $S_k$ computes the truck's estimated waiting time $\Tilde{w}_k$ according to the present charging schedule and sends it to the truck.
    \item[3)] The truck and its distant stations $S_{\ell}$, $\ell\!=\!k+1,\dots,N$ go through two rounds of communications and computations. In the end, the truck receives the estimated waiting times $\Tilde{w}_{\ell}$, $\ell\!=\!k\!+\!1,\dots,N$ from these stations.
    \item[4)] Based on the estimated waiting times $\Tilde{w}_k,\Tilde{w}_{k+1},\dots,\Tilde{w}_N$, the truck updates its charging plan and sends to station $S_k$ its charging decision $(b_k^*,t_k^*)$.
    \item[5)] Upon receiving the charging decision $(b_k^*,t_k^*)$, station $S_k$ places the truck into the queue and updates related information if $b_k^*$ is one.
\end{itemize}

A summary of the communication and coordination architecture is depicted in Figure~\ref{Fig.2}. Note that upon a truck's arrival at its $k$th ramp $r_k$, there are communications between the truck and all the subsequent stations $S_k,S_{k+1},\dots,S_N$ along the truck's pre-planned route. However, only station $S_k$ receives the updated charging plan and adjusts the schedule accordingly. This leads to limited computation related to the scheduling for the stations, as we will show in the next section. 

The two rounds of communications and computations mentioned above are essential for the estimation of waiting times at distant stations, which builds upon the estimated arrival-time windows provided by the truck. More specifically, the truck first provides the distant stations $S_{\ell}$, $\ell\!=\!k\!+\!1,\dots,N$ with its estimated earliest arrival times. These times enable stations to respond with estimated maximum waiting times $\Bar{w}_{k+1},\dots,\Bar{w}_N$. Using this information, the truck computes its estimated latest arrival times to these stations and sends them back to the stations. The truck's estimated arrival-time window at each station is defined by its earliest and latest arrival times, which are leveraged by the stations $S_{\ell}$, $\ell\!=\!k+1,\dots,N$ to compute the estimated waiting times $\tilde{w}_{k+1},\dots,\tilde{w}_N$ for the truck. The computations at stations rely on their respective waiting time forecast models. 

Note that in the proposed coordination framework, there is no communication between stations or between individual trucks; communication occurs only between trucks and stations. Moreover, all the computations are carried out locally, so the coordination framework is fully distributed. In what follows, we provide details on the computations conducted by the stations and trucks, respectively.

\section{Estimation and Computation at Charging Stations}\label{Section III}
This section presents the estimation and computation carried out at charging stations. We start by discussing the process for stations to establish local waiting time forecast models. We then introduce how stations calculate the estimated waiting times upon trucks' requests utilizing the obtained forecast models. Depending on how far the trucks are to the stations when sending out the requests, the stations compute the trucks' waiting times either exactly based on the present charging schedule, or approximately based on the forecast models. In the end, we discuss the approach each station takes for charging scheduling and information updates.

\subsection{Waiting Time Forecast Model}\label{Subsection III.A}
As discussed earlier, stations operate in two different phases: the data collection phase and the nominal operation phase. During the first phase, charging stations record historical data of trucks that used the stations for charging over a fixed period. This data includes the arrival times of the trucks, the port each truck was assigned to, and the trucks' actual waiting times at the station. Using this information, each charging station computes its waiting time forecast model to facilitate the waiting time estimations, as will be introduced in Section~\ref{Subsection III.C}. 

In particular, the forecast model can be viewed as a function $f$ that outputs a nonnegative waiting time estimation given an estimated arrival time of a truck. In this paper, the forecast model at each station is attained by discretizing the timeline of one day, and then computing the average waiting times of all trucks documented in the historical data within each time interval. Such a modeling method is simple and reliable, as indicated in our results. Other approaches to establishing the forecast model based on historical data at local stations may also fit into the proposed coordination framework.  

\subsection{Waiting Time Computation for Nearby Trucks}
Each station classifies trucks that communicate with it into two types: \emph{nearby trucks} and \emph{distant trucks}. A station views a truck as nearby if it has reached the corresponding ramp connecting to the station. In such a case, the station can compute the trucks' waiting time estimate according to the present schedule, thus making the accurate estimation.

More specifically, given a charging station, let us denote by $C$ the collection of its charging ports. Each charging port $c\!\in\! C$ maintains an earliest available time $t_{a,c}$, which represents the earliest time since when the port becomes available onwards indefinitely. If a nearby truck reaches the corresponding ramp of the station, it sends its arrival time $a\!+\!d$ to the station, where $a$ denotes the present time (at which the truck reaches the ramp), and $d$ denotes the detour time required by the truck to reach the station from the ramp. The station then computes the truck's waiting time at the station by 
\begin{align}
\tilde{w}=\max\Big\{\min_{c\in{C}}\big(t_{a,c}-(a+d)\big),0\Big\},\label{Eq.1}
\end{align}
which compares the available times of every port at the station with the truck's estimated arrival time and calculates the waiting time accordingly. This waiting time is then sent to the truck used for its charging plan optimization.   

\subsection{Waiting Time Estimation for Distant Trucks}\label{Subsection III.C}
Compared to waiting time computations for nearby trucks, it is more challenging for a station to calculate the waiting time estimations for distant trucks. This difficulty arises because the estimated waiting time of a distant truck is determined by its arrival time at the station, which is coupled with the truck's charging decisions and actual waiting times at all preceding stations. Our framework addresses this difficulty by decoupling estimation and computation using an estimated arrival time computed by the truck and performing two rounds of communications and computations between the station and the distant truck, as introduced below.

We denote by $\underline{a}\!+\!d$ the estimated earliest arrival time sent from a distant truck to the station. Upon receiving this information, the station computes the maximum waiting time $\Bar{w}$ since $\underline{a}\!+\!d$ based on its waiting time forecast model, i.e.,
\begin{align}
\Bar{w}=\max_{a\geq\underline{a}}f(a+d).\label{Eq.2}
\end{align}
The station then replies to the truck with $\Bar{w}$. As a response, the truck provides its estimated latest arrival time $\bar{a}$ to the station. The procedures for computing $\underline{a}$ and $\bar{a}$ will be detailed in Section~\ref{Section IV}. Through the communication, the station receives an estimated arrival-time window from the distant truck, represented as the closed interval $[\underline{a}+d,\Bar{a}+d]$. Based on this arrival-time window, the station can compute the estimated waiting time $\tilde{w}$ for the distant truck by
\begin{align}
\tilde{w}=\frac{1}{\bar{a}-\underline{a}}\int_{\underline{a}+d}^{\bar{a}+d}f(t)\,dt,\label{Eq.3}
\end{align}
where the waiting time forecast model $f$ is established by the station locally. In response to the truck's request, the station sends the estimated waiting time $\tilde{w}$ to the truck; cf. Figure~\ref{Fig.2}.

\subsection{Charging Scheduling and Information Update}
In the proposed coordination framework, each station computes and provides waiting time estimates $\Tilde{w}$ to both nearby and distant trucks, but only assigns charging ports for nearby trucks upon receiving their charging plans, as illustrated in Figure~\ref{Fig.2}. Such a design simplifies the charging scheduling computations at stations. More importantly, it enables trucks to adjust their charging plans at each preceding ramp to address travel uncertainties, such as travel time delays, charging congestion at stations, etc. 

To proceed, we introduce how stations assign their charging ports to individual trucks following the first-come, first-served rule, and update the charging schedule information accordingly. A station receives the charging plan from a nearby truck in the form of its arrival time $a\!+\!d$ and charging decision $(b^*,t^*)$, where $b^*$ stands for the binary charging decision of the truck, and $t^*$ denotes the planned charging time at the station. If $b^*\!=\!0$, the earliest availability times $\{t_{a,c}\}_{c\in{C}}$ remain unchanged for all the charging ports $c$. Otherwise, let us denote by $c^*$ the port that achieves the minimal waiting time in \eqref{Eq.1}, equivalently,
\begin{align}
c^*\in\arg\min_{c\in{C}}t_{a,c}.\label{Eq.4}
\end{align}
The port $c^*$ is assigned to the nearby truck for a duration of $t^*$, starting from $a+d+\tilde{w}$, where $\Tilde{w}$ is given by~\eqref{Eq.1}. Accordingly, the earliest available time of the port $c^*$ is updated as 
\begin{align}
t_{a,c^*}=a+d+\tilde{w}+t^*,\label{Eq.5}
\end{align}
while the available times of other ports remain unchanged. 

Note that the mechanism introduced above follows the first-come, first-served rule. It ensures that charging time is spent consecutively at the same port and is straightforward to implement. In addition, we assume that the station communicates with only one nearby truck at a time for simplification of the discussion. The proposed framework also allows for alternative scheduling approaches at the potential expense of more communication and computation at the stations. 

\section{Charging Strategy Optimization by Trucks}\label{Section IV}
This section presents the charging strategy optimization problem addressed by the trucks, building upon the estimated waiting times provided by the charging stations along their pre-planned routes. We start by introducing the route models, decision variables for individual trucks, and the dynamic models used to formulate the charging planning problem. Next, we describe how trucks compute their estimated arrival-time windows for requesting waiting time estimates. Finally, we propose a method for determining the optimal charging strategy for the truck. 

\subsection{Model of the Charging Planning Problem} \label{Subsection IV.A}
\begin{figure}[t]
     \centering
     \includegraphics[width=0.95\linewidth]{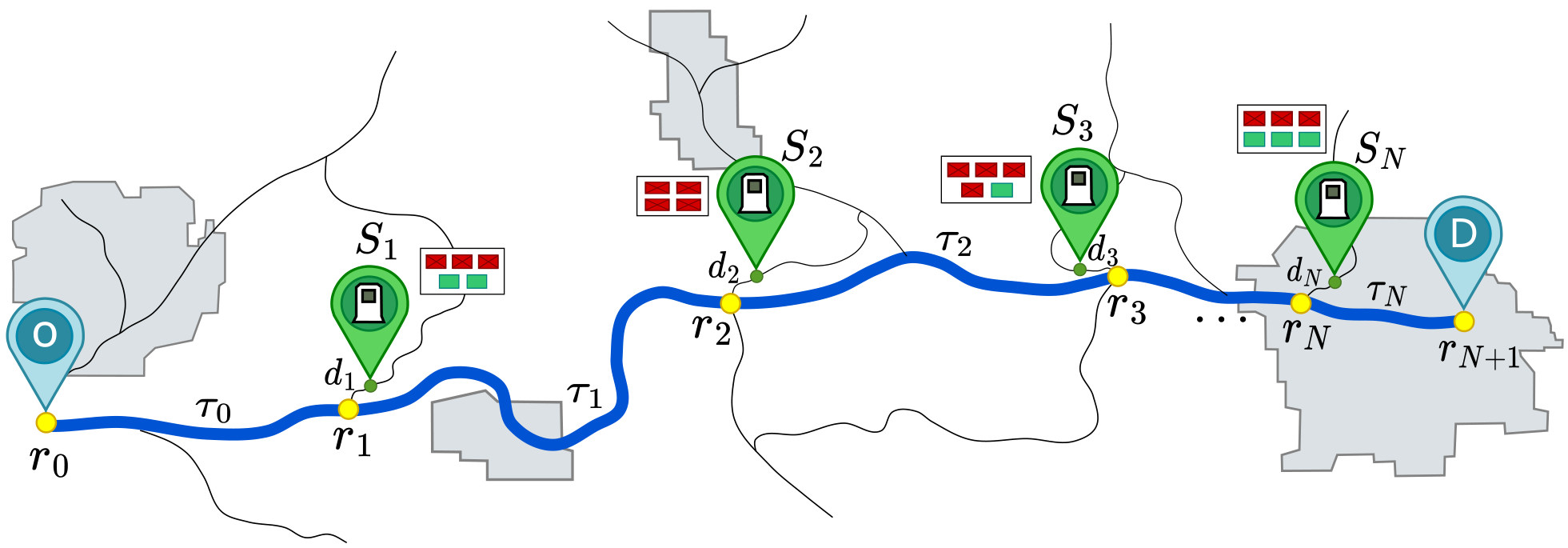}
     \vspace{-8pt}
      \caption{The route model of each truck, where the pre-planned route between the origin and destination is represented by the blue path. Ramps in the route are marked in yellow and the connected charging stations $S_k$, $k\!=\!1,\dots, N$ are shown by green labels. Green and red blocks at each station indicate the availability of the charging ports.}
      \label{Fig.3}
   \end{figure}
We assume each truck has a fixed, pre-planned route between its origin and destination, as illustrated in Figure~\ref{Fig.3}. For a given truck, there are $N$ charging stations $S_k$, $k\!=\!1,\dots,N$ along its route, each connected to a corresponding ramp $r_k$. In particular, the origin and destination are denoted by $r_0$ and $r_{N+1}$, respectively. The pre-planned route is divided into $N\!+\!1$ road segments by the ramps, with the nominal traverse time from $r_k$ to $r_{k+1}$ denoted as $\tau_k$, $k\!=\!0,\dots,N$. The actual travel times on the main route are subject to bounded disturbances. Specifically, the actual travel time from $r_k$ to $r_{k+1}$ is denoted as $\tau_k\!+\!\delta_{\tau,k}$, where $\delta_{\tau,k}$ is a zero-mean random variable generated within the bounded range $[-\delta_k^1 \tau_k,\delta_k^1 \tau_k]$. Here, $\delta_k^1\!\in\!\Re_+$, $k\!=\!0,\dots,N$, are known constants. The detour time between $r_k$ and $S_k$ is denoted by $d_k$, which is relatively short and assumed to be deterministic in our model. 

Trucks optimize their charging plans dynamically each time they arrive at a ramp $r_k$, $k\!=\!1,\dots,N$, aiming to minimize the total operational costs required to traverse the remaining routes for completing their delivery tasks. As discussed in Section~\ref{Section II}, each truck makes the following two types of decisions upon reaching a ramp $r_k$: (i) whether to charge at $S_{\ell}$, $\ell\!=\!k,\dots,N$; and (ii) how long to charge the truck if it decides to charge at $S_{\ell}$. These decisions can be represented by the variables
\begin{equation}
    b_{\ell}\!\in\!\{0,1\},\;t_{\ell}\!\in\!\Re_+,\quad  \ell\!=\!k,\dots,N,\label{Eq.6}
\end{equation}
where $b_{\ell}\!=\!1$ if the truck decides to charge at the $k$th charging station along its route and $0$ otherwise, $t_{\ell}$ denotes the planned charging time of the truck at $S_{\ell}$ if $b_{\ell}\!=\!1$.

To describe the charging plan, it is necessary to model the remaining batteries upon reaching different ramps, which we introduce next. For any given truck, let $e_{\text{ini}}$ be the initial battery of the truck at its origin and $e_k$ its remaining battery when it first arrives at its ramp $r_k$. Similar to travel times $\tau_k$, we assume there are random fluctuations $\delta_{e,k}$ of energy consumption while the truck travels between ramps $r_k$ and $r_{k+1}$, where $\delta_{e,k}$, $k\!=\!0,\dots,N$, take values from the intervals $[-\delta_k^2\bar{P}\tau_k,\delta_k^2 \bar{P}\tau_k]$ and $\delta_k^2\!\in\!{\Re_+}$ are known constants. As a result, the dynamic of the remaining battery of a truck upon reaching its ramp $r_1$ is modeled as
\begin{equation}
    e_1=e_{\text{ini}}-\bar{P}\tau_0+\delta_{e,0},\label{Eq.7}
\end{equation}
while for $\ell\!=\!k,\dots,N$ with $k\!=\!1,\dots,N$, the values $e_{\ell+1}$ are given by
\begin{equation}
    e_{\ell+1}=e_{\ell}+b_{\ell}\Delta{e_{\ell}}-\bar{P}(2b_{\ell}d_{\ell}+\tau_{\ell})+\delta_{e,\ell}.\label{Eq.8}
\end{equation}
Here $\bar{P}$ denotes the truck's battery consumption per unit of travel time, and $\Delta{e}_{\ell}$ denotes the increased battery charged at $S_{\ell}$. To ensure that the truck's remaining battery at $r_{\ell}$ is sufficient for reaching $S_{\ell}$, we require that
\begin{subequations}
\label{Eq.9}
\begin{align}
&e_{\ell}\geq{e_s}+\bar{P}d_{\ell}, \quad \ell\!=\!k,\dots,N,\label{Eq.9a}\\
&e_{N+1}\geq{e_s},\label{Eq.9b}
\end{align}
\end{subequations} 
where $e_s$ is a constant margin on the remaining battery for safe operation. In \eqref{Eq.8}, the charging process at each station is approximated by a linear model, given as
\begin{equation}
    \Delta{e_{\ell}}=t_{\ell}\min\big\{P_{\ell},P_{\max}\big\},\quad \ell\!=\!k,\dots,N,\label{Eq.10}
\end{equation}
where $P_{\ell}$ denotes the charging power provided by $S_{\ell}$ while $P_{\max}$ is the maximum charging power acceptable by the truck's battery. Restricted by the battery capacity of the truck, $\Delta{e_{\ell}}$ is confined by  
\begin{equation}
    {0}\leq\Delta{e_{\ell}}\leq{e_f}\!-\!(e_{\ell}\!-\!\bar{P}d_{\ell}),\quad \ell\!=\!k,\dots,N,\label{Eq.11}
\end{equation}
where $e_f$ is the full battery of the truck.

To effectively coordinate trucks' arrival times at the stations, we also need to model the predicted arrival times of trucks at their ramps. Let $a_{\text{ini}}$ denote the departure time of a truck from its origin, and $a_{\ell}$ the time it reaches ramp $r_{\ell}$ for the first time. The arrival time of the truck at its first ramp can be then described as
\begin{equation}
    a_1=a_{\text{ini}}+\tau_0+\delta_{\tau,0}.\label{Eq.12}
\end{equation}
The arrival times $a_{\ell+1}$ for $\ell\!=\!k,\dots,N$ with $k\!=\!1,\dots,N$, are given by
\begin{equation}
    a_{\ell+1}=a_{\ell}+b_{\ell}(t_{\ell}\!+\!w_{\ell}\!+\!2d_{\ell})+\tau_{\ell}+\delta_{\tau,\ell},\label{Eq.13}
\end{equation}
where $w_{\ell}\!\in\!\Re_{+}$ represents the truck's waiting time at $S_{\ell}$ due to charging congestion. Typically, delivery missions have deadlines $a_{\text{end}}$, which can be described as the constraint $a_{N+1}\leq a_{\text{end}}$. However, due to constraints of charging resources, it is difficult to ensure the deadline as a hard constraint. We encode the preference of reaching the destination before the deadline as a soft constraint, as we will discuss next in Section~\ref{Subsection IV.C}.

We incorporate bounded uncertainties in travel times and battery consumption into the proposed coordination framework. In practice, there may be intricate relationships between the two uncertain disturbances. For instance, a special case of $\delta_{e,\ell}$ in (\ref{Eq.8}) can be represented by $\Bar{P}\delta_{\tau,\ell}$. To suit more general scenarios, we treat the two uncertainties as random variables and leave their relationship unspecified.

\subsection{Computation of Estimated Arrival-Time Windows} \label{Subsection IV.B}
A truck computes its charging plan upon reaching each ramp. When arriving at the $k$th ramp $r_k$, the truck optimizes its charging plan, including its charging decisions at every subsequent station $S_{\ell}$, $\ell\!=\!k,\dots,N$, along its remaining route, and the target is to complete the delivery mission while minimizing its operational costs. To effectively coordinate the truck's arrival and charging times at these stations, the estimated waiting times $\Tilde{w}_k,\dots,\Tilde{w}_N$ provided by the corresponding stations are required.

As discussed in Section~\ref{Section II}, these estimated waiting times rely on the truck's arrival times at the stations. Upon reaching the $k$th ramp 
$r_k$, the truck provides its exact arrival time $a_k\!+\!d_k$ to the \emph{nearby station} $S_k$. This enables station $S_k$ to compute the estimated waiting time $\tilde{w}_k$ according to the real-time charging schedule; cf. \eqref{Eq.1}. For the \emph{distant stations} $S_{\ell}$, $\ell\!=\!k\!+\!1,\dots,N$, however, it is not possible for the truck to provide its accurate arrival times. This is because each of the truck's arrival times $a_{\ell}\!+\!d_{\ell}$, $\ell=k+1,\dots,N$, is coupled with its charging decisions and the actual waiting times at all the preceding stations $S_k,\dots,S_{\ell-1}$. The situation is further exacerbated by the presence of travel uncertainties.

To address this challenge, the truck computes estimated arrival-time windows for each distant station and sends the information to these stations. In particular, upon reaching ramp $r_k$, the estimated arrival-time windows for its distant stations $S_{\ell}$, $\ell\!=\!k+1,\dots,N$ are given by the closed intervals $[\underline{a}_{\ell}+d_{\ell},\bar{a}_{\ell}+d_{\ell}]$. Here, $\underline{a}_{\ell}$ and $\bar{a}_{\ell}$ represent the estimated earliest and latest arrival times of the truck at ramp $r_{\ell}$,  respectively, while $d_{\ell}$ is the detour time to reach $S_{\ell}$ from the corresponding ramp. In what follows, we present how the truck computes $\underline{a}_{\ell}$ and $\bar{a}_{\ell}$ for the distant stations $S_{\ell}$, $\ell\!=\!k+1,\dots,N$. 

\subsubsection{The Estimated Earliest Arrival Time} The estimated earliest arrival time $\underline{a}_{\ell}$ is computed by assuming the best scenarios, where the charging amount is only sufficient to reach $S_{\ell}$, the waiting times at all the stations preceding to $S_{\ell}$ are assumed to be zero, and all the random variables are assumed to take favorable values. Specifically, when a truck reaches its ramp $r_k$ at the time instant $a_k$ with the remaining battery $e_k$, it solves $N-k$ optimization problems, each corresponding to a distant station. For $\ell=k+1,\dots,N$, the optimization problem addressed by the truck is  
\begin{equation}
    \min_{\{(b_h,t_h)\}_{h=k}^{\ell-1}}\sum_{h=k}^{\ell-1}b_h\big(2d_h\!+\!t_h\big),\label{Eq.14}
\end{equation}
subject to the constraints 
\begin{subequations}
\label{Eq.15}
\begin{align}
& b_h\!\in\!\{0,1\},~t_h\!\in\!\Re_{+},\label{Eq.15a}\\
& e_{h+1}=e_h\!+\!b_h\Delta{e}_h\!-\!\bar{P}(2b_hd_h\!+\!\tau_h)\!+\!\delta_h^2\bar{P}\tau_h,\label{Eq.15b}\\
& e_{h+1}\geq{e_s\!+\!\bar{P}d_{h+1}},\label{Eq.15c}\\
&\Delta{e}_h=t_h\min\{P_h,P_{\max}\},\label{Eq.15d}\\
& 0\leq{\Delta{e_h}}\leq{e_f\!-\!(e_h\!-\!\bar{P}d_h)},\label{Eq.15e}
\end{align}
\end{subequations}
for $h=k,\dots,\ell-1$. Supposing that the minimum is attained at $\{(\hat{b}_h,\hat{t}_h)\}_{h=k}^{\ell-1}$, the estimated earliest arrival times of the truck at the distant ramp $r_{\ell}$ is then computed by
\begin{align}
\underline{a}_{\ell}=a_k+\sum_{h=k}^{\ell-1}\Big(\hat{b}_h(2d_h\!+\!\hat{t}_h)\!+\!\tau_h\!-\!\delta_h^1{\tau_h}\Big).\label{Eq.16}
\end{align}
In computing the truck's earliest arrival times at distant stations, the favorable uncertainties in the energy consumption model in \eqref{Eq.15b} taking the values $\delta_h^2e_h$ and those in the travel times taking the values of $-\delta_h^1\tau_h$, as given in \eqref{Eq.16}. The estimated earliest arrival times to $\underline{a}_{\ell}\!+\!d_{\ell}$ are then sent to the distant stations $S_\ell$, $\ell\!=\!k\!+\!1,\dots,N$.   

\subsubsection{The Estimated Latest Arrival Time} The latest arrival times are obtained by assuming the opposite, i.e., fully charging at all preceding stations, long waiting times, and unfavorable values of random variables. The computation of the latest arrival times does not require solving optimization problems. Instead, they can be derived from closed-form formulas by employing values of the variables in the worst scenarios. First, to attain the long estimated waiting times at stations, the truck relies on the response of the distant stations to the estimated earliest arrival times. More specifically, the truck arriving at its ramp $r_k$ first sends its earliest arrival times $\underline{a}_{\ell}\!+\!d_{\ell}$ to the distant stations $S_{\ell}$. In response, the truck receives the estimated maximum waiting time $\bar{w}_\ell$ from $S_{\ell}$, $\ell\!=\!k\!+\!1,\dots,N$; cf. \eqref{Eq.2}. According to these estimated maximum waiting times, the truck can compute its latest arrival times at the corresponding ramps leading to these stations by
\begin{align}
\!\!\!\!\bar{a}_{\ell}\!=\!a_k\!+\!\sum_{h=k}^{\ell-1}\!\Big(2d_h\!+\!\frac{\Delta \Bar{e}_h}{\min\{P_h,P_{\max}\}}\!+\!\bar{w}_h\!+\!\tau_h\!+\!\delta_h^1\tau_h\Big),\label{Eq.17}
\end{align}
where, the estimated maximum waiting time at the nearby station $\Bar{w}_k$ equals the estimated waiting time $\tilde{w}_k$. As the battery is fully charged at every station preceding to $S_{\ell}$, in line with (\ref{Eq.10}), the maximum increased battery $\Delta \bar{e}_k$ is obtained by
\begin{align}
\Delta \Bar{e}_k=e_f-(e_k-\Bar{P}d_k),\label{Eq.18}
\end{align}
while for $h\!=\!k\!+\!1,\dots,\ell\!-\!1$, there is
\begin{align}
\Delta \Bar{e}_h=\bar{P}(d_{h-1}\!+\!\tau_{h-1}\!+\!d_h)+\delta_{h-1}^2\bar{P}\tau_{h-1}.\label{Eq.19}
\end{align}
As such, the truck computes its estimated arrival-time windows $[\underline{a}_{\ell}+d_{\ell},\bar{a}_{\ell}+d_{\ell}]$, $\ell=k+1,\dots,N$ at every distant station $S_{\ell}$ along its remaining route. By sending this information to $S_{\ell}$, the stations can provide the estimated waiting times $\tilde{w}_{\ell}$; cf.~\eqref{Eq.3}. Together with the waiting time $\Tilde{w}_k$ provided by the nearby station $S_k$, the truck can update its charging strategy.

For the estimated arrival-time windows, we have the following result holds. 
\begin{proposition}\label{Proposition 1}
    Let $\underline{a}_\ell^k$ and $\bar{a}_\ell^k$ be the earliest and latest arrival times of a truck at ramp $r_{\ell}$ computed at ramp $r_k$. Suppose that the truck's effective charging powers at stations $S_k$ and $S_{k+1}$ are identical, i.e., $\min\{P_k,P_{\max}\}=\min\{P_{k+1},P_{\max}\}$, and the actual waiting time at station $S_k$ is upper-bounded by $\bar{w}_k$. Then for $\ell\!=\!k\!+\!2,\dots,N$, we have
    \begin{subequations}
    \label{Eq.20}
    \begin{align}
    &\underline{a}_\ell^k\leq \underline{a}_\ell^{k+1},\label{Eq.20a}\\
    &\bar{a}_\ell^k\geq \bar{a}_\ell^{k+1}.\label{Eq.20b}
    \end{align}
    \end{subequations}
\end{proposition}
\begin{proof} 
See the Appendix.
\end{proof}

\begin{remark}\label{Remark1}
    Proposition~\ref{Proposition 1} states that as a truck travels from ramp $r_k$ to ramp $r_{k+1}$, its estimated arrival-time windows at all the distant stations $S_{\ell}$, for $\ell\!=\!k\!+\!2,\dots,N$, shrink. This facilitates more accurate waiting time estimates at each station $S_{\ell}$. Based on this result, we will assume in Theorem~\ref{Theorem 1} that the accuracy of the waiting time estimates provided by each distant station $S_{\ell}$ improves as the truck approaches the station. 
\end{remark}

\subsection{Charging Strategy Optimization}\label{Subsection IV.C}
The preceding subsection presents the detailed computation by the truck to facilitate the calculation of the estimated waiting times $\Tilde{w}_k,\dots,\Tilde{w}_N$ by the corresponding stations. Once receiving the estimated waiting times, the truck optimizes its charging plan by solving the problem
\begin{align}
\min_{\{(b_\ell,t_\ell)\}_{\ell=k}^N}\;~J_k=&\sum_{\ell=k}^N \Big(\xi{b_{\ell}}\big(2d_\ell\!+\!t_\ell\!+\!\tilde{w}_\ell\big)+\theta_{\ell}b_{\ell}t_{\ell}\Big)+\nonumber\\&\max\Big\{\gamma(a_{N+1}-a_{\text{end}}),0\Big\},\label{Eq.21}
\end{align}
subject to the constraints
\begin{subequations}
\begin{align}
& b_\ell\!\in\!\{0,1\},~t_\ell\!\in\!\Re_{+},\label{Eq.22a}\\
& e_{\ell+1}=e_\ell+b_\ell\Delta{e}_\ell-\bar{P}(2b_\ell d_\ell+\tau_\ell),\label{Eq.22b}\\
&\Delta{e}_\ell=t_\ell\min\{P_\ell,P_{\max}\},\label{Eq.22c}\\
& 0\leq{\Delta{e_\ell}}\leq{e_f\!-\!(e_\ell\!-\!\bar{P}d_\ell)},\label{Eq.22d}\\
& a_{\ell+1}=a_\ell+b_\ell(t_\ell+\tilde{w}_\ell+2d_\ell)+\tau_\ell,\label{Eq.22e}
\end{align}
\end{subequations}
for $\ell=k,\dots,N$, and the constraints
\begin{subequations}
\label{Eq.23}
\begin{align}
& e_{k+1}\geq{e_s+\delta_{k}^2\bar{P}\tau_{k}+\!\bar{P}d_{k+1}},\label{Eq.23a}\\
& e_{\ell}\geq{e_s+\!\bar{P}d_{\ell}}, \quad \ell\!=\!k\!+\!2,\dots,N,\label{Eq.23b}\\
& e_{N+1}\geq{e_s}.\label{Eq.23c}
\end{align}
\end{subequations}

The cost function is composed of labor costs, charging expenses, and a penalty for violating the delivery deadline. Here $\xi$ denotes the labor cost per time unit incurred due to detours, charging, and waiting, while $\theta_{\ell}$ represents the electricity cost at the station $S_{\ell}$ per unit of charging time. Additionally, $\gamma$ serves as a tuning parameter used to encode the delivery deadline as a soft constraint, with $a_{\text{end}}$ denoting the latest arrival time indicated by the deadline.

\begin{figure*}[t]
     \centering
     \includegraphics[width=0.85\linewidth]{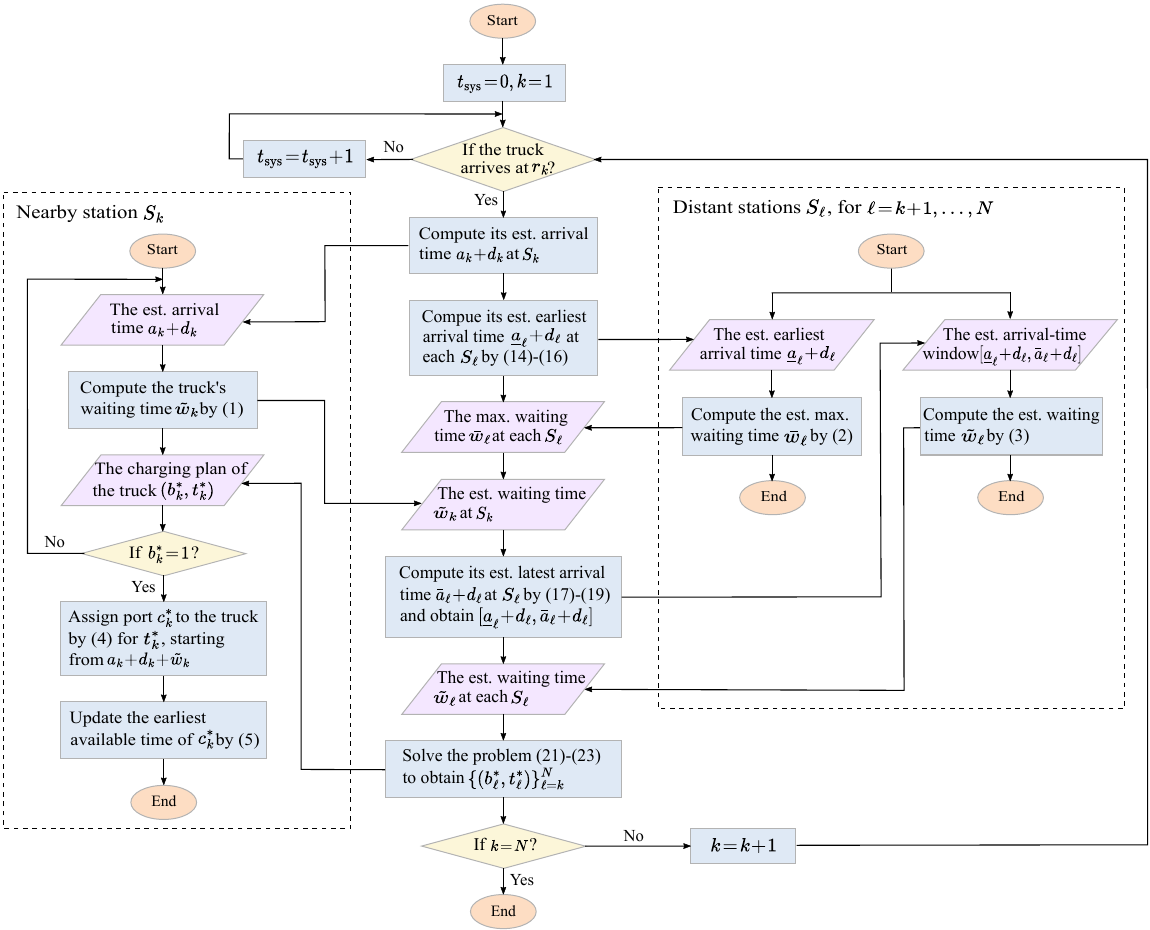}
     \vspace{-5pt}
      \caption{A flow chart illustrating the operation procedure of the coordination framework, taken from one truck's perspective. Here, est. stands for estimated, max. stands for maximum, and $t_{sys}$ represents the system time used to monitor the status of every truck.}
      \label{Fig.4}
   \end{figure*}

The constraints \eqref{Eq.22b} and \eqref{Eq.22e} are obtained via certainty equivalence approximation of \eqref{Eq.8} and \eqref{Eq.13}, i.e., by setting the random variables $\delta_{e,\ell}$ and $\delta_{\tau,\ell}$ to their mean values. The safety margin in \eqref{Eq.23a} is increased by $\delta_k^2\bar{P}\tau_k$ when compared with~\eqref{Eq.9}. This ensures the feasibility of the obtained solution at $S_k$ under all possible values of the random variables. Suppose that the optimal value is attained at $\{(b_{\ell}^*,t_{\ell}^*)\}_{\ell=k}^N$, the truck sends to the nearby station $S_k$ its decision $(b_k^*,t_k^*)$ and charges at the station for the time duration of $t_k^*$ if $b_k^*\!=\!1$. For better understanding, a flow chart that illustrates the operation of the proposed coordination framework is provided in Figure~\ref{Fig.4}.
   
\begin{remark}\label{Remark2}
The problem \eqref{Eq.21}-\eqref{Eq.23} is nonconvex and represents a mixed-integer program with bilinear constraints, which cannot be directly addressed by many standard solvers. We first linearize the bilinear constraints, leading to a mixed-integer linear program (MILP). Then we solve the resulting MILP using Gurobi. We refer readers to \cite[Appendix F]{bai2023} for details of the linearization procedure of a similar problem. While the worst-case complexity of the MILP grows exponentially with $N$, such a case did not occur in our computational studies. For large values of $N$, the original problem can be addressed efficiently by a rollout-based approach studied in~\cite{bai2023}, which provides a near-optimal solution to the problem by solving at most $N\!+\!1$ linear programs.
\end{remark}

\begin{remark}\label{Remark3}
The optimal charging plan $(b_{k}^*,t_{k}^*)$ computed at each ramp $r_k$, $k\!=\!1,\dots,N$, is feasible, i.e., regardless of the values of the random variables $\delta_{e,k}$, the remaining battery of the truck when it reaches ramp $r_{k+1}$, denoted as $\hat{e}_{k+1}$, satisfies $\hat{e}_{k+1}\!\geq\! e_s\!+\!\Bar{P}d_{k+1}$. This is guaranteed by \eqref{Eq.23a}.
\end{remark}

For the proposed charging strategy, we have the following results hold.

\begin{theorem}\label{Theorem 1}
Let $J_k^*$ be the optimal cost function of \eqref{Eq.21} computed at ramp $r_k$ attained at $\{(b_{\ell}^*,t_{\ell}^*)\}_{\ell=k}^N$. Suppose that $\tilde{w}_\ell^k$ is the estimated waiting time at station $S_\ell$ when the truck reaches ramp $r_k$, and $\hat{w}_\ell^k$ is the actual waiting time at $S_{\ell}$ by applying the charging plan $\{(b_h^*,t_h^*)\}_{h=k}^{\ell-1}$. Let $\hat{J}_k$ be the cost function of \eqref{Eq.21} by following $\{(b_{\ell}^*,t_{\ell}^*)\}_{\ell=k}^N$ and replacing $\tilde{w}_\ell^k$ with $\hat{w}_\ell^k$, $\ell=k,\dots,N$. 
\begin{itemize}
\item [(a)] For $k\!=\!1,\dots,N\!-\!1$, $|\hat{J}_k-J_k^*|$ has the upper bound $\Delta{\bar{J}}_k$ of the form
\begin{align}
\Delta{\bar{J}}_k=(\xi\!+\!\gamma)\!\!\sum_{\ell=k+1}^N\!\!\Delta{w}_{\ell}^k, \label{Eq.24}
\end{align}
where $\Delta{w}_{\ell}^k\!=\!|\hat{w}_{\ell}^k\!-\!\tilde{w}_{\ell}^k|$. In particular, $\Delta{\bar{J}}_N\!=\!0$. 
\item [(b)] Suppose that $\Delta{w}_{\ell}^{k+1}\!\leq\!{\Delta{w}_{\ell}^k}$ for $\ell\!=\!k\!+\!2,\dots,N$. For $k\!=\!1,\dots,N\!-\!1$, we have
\begin{align}
\Delta{\bar{J}}_{k+1}-\Delta{\bar{J}}_k\leq{0}.\label{Eq.25}
\end{align}
\end{itemize}
\end{theorem}
\begin{proof} 
See the Appendix.
\end{proof}

Here $\Delta w_\ell^k$ is used to quantify the waiting time estimate error at station $S_\ell$ when the truck reaches ramp $r_k$ with $k\!<\! \ell$. Intuitively, the condition $\Delta{w}_{\ell}^{k+1}\!\leq\!{\Delta{w}_{\ell}^k}$ can be viewed as a consequence of improved accuracy of the arrival time estimate, as shown in Proposition~1; cf. Remark~\ref{Remark1}. Theorem~\ref{Theorem 1} indicates that $|\hat{J}_k\!-\!J_k^*|$ at each computation instant $k$ is upper bounded by $\Delta{\bar{J}}_k$. Moreover, such a bound decreases progressively as the charging plan updates.

\section{Simulation Studies over the Swedish Road Network}\label{Section V}
\begin{figure*}[t]
\centering
\begin{minipage}{1\textwidth}
\centering
\subfigure[Road transport flow]
{\includegraphics[width=0.32\textwidth]{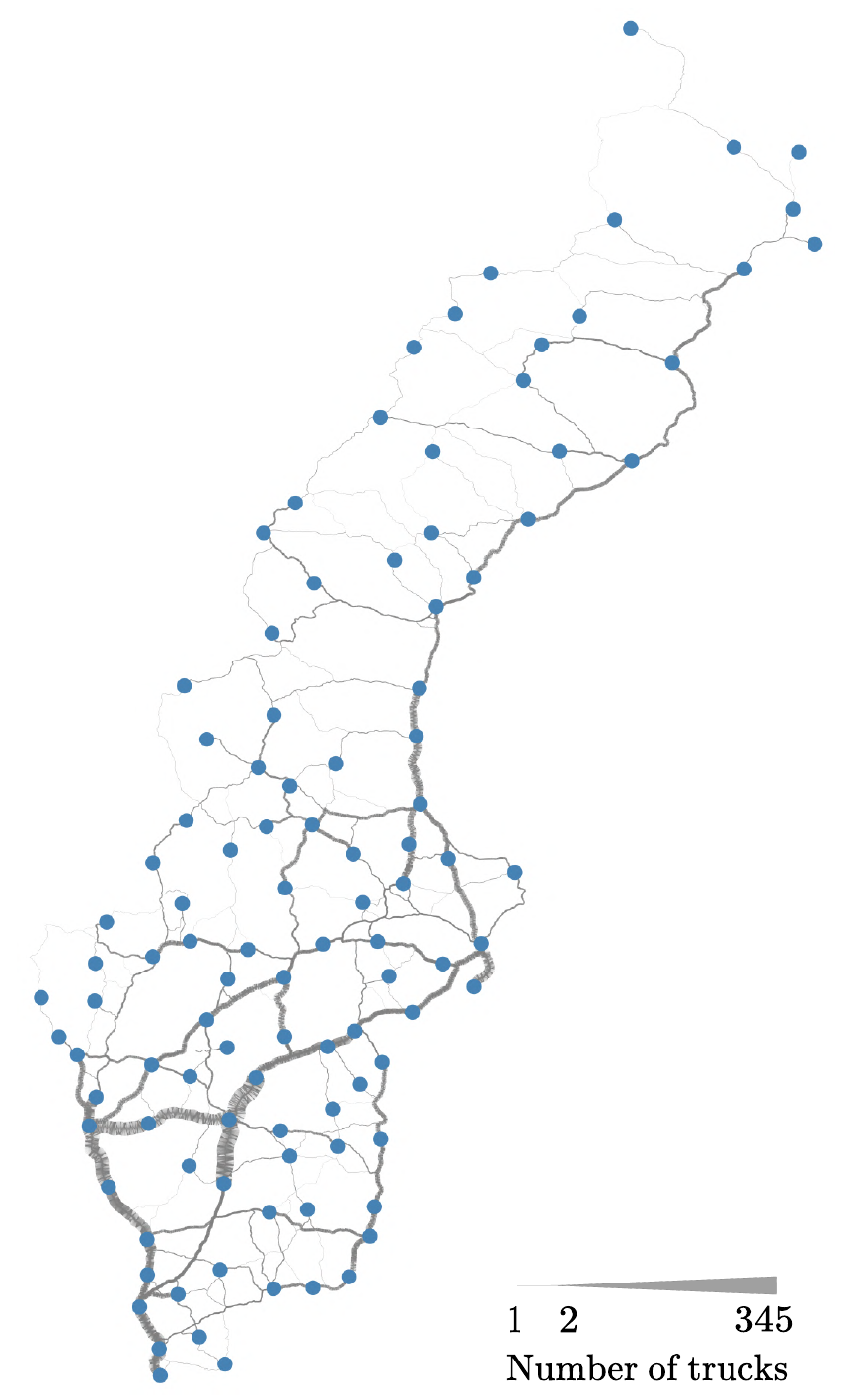}}
\subfigure[Potential charging stations]
{\includegraphics[width=0.32\textwidth]{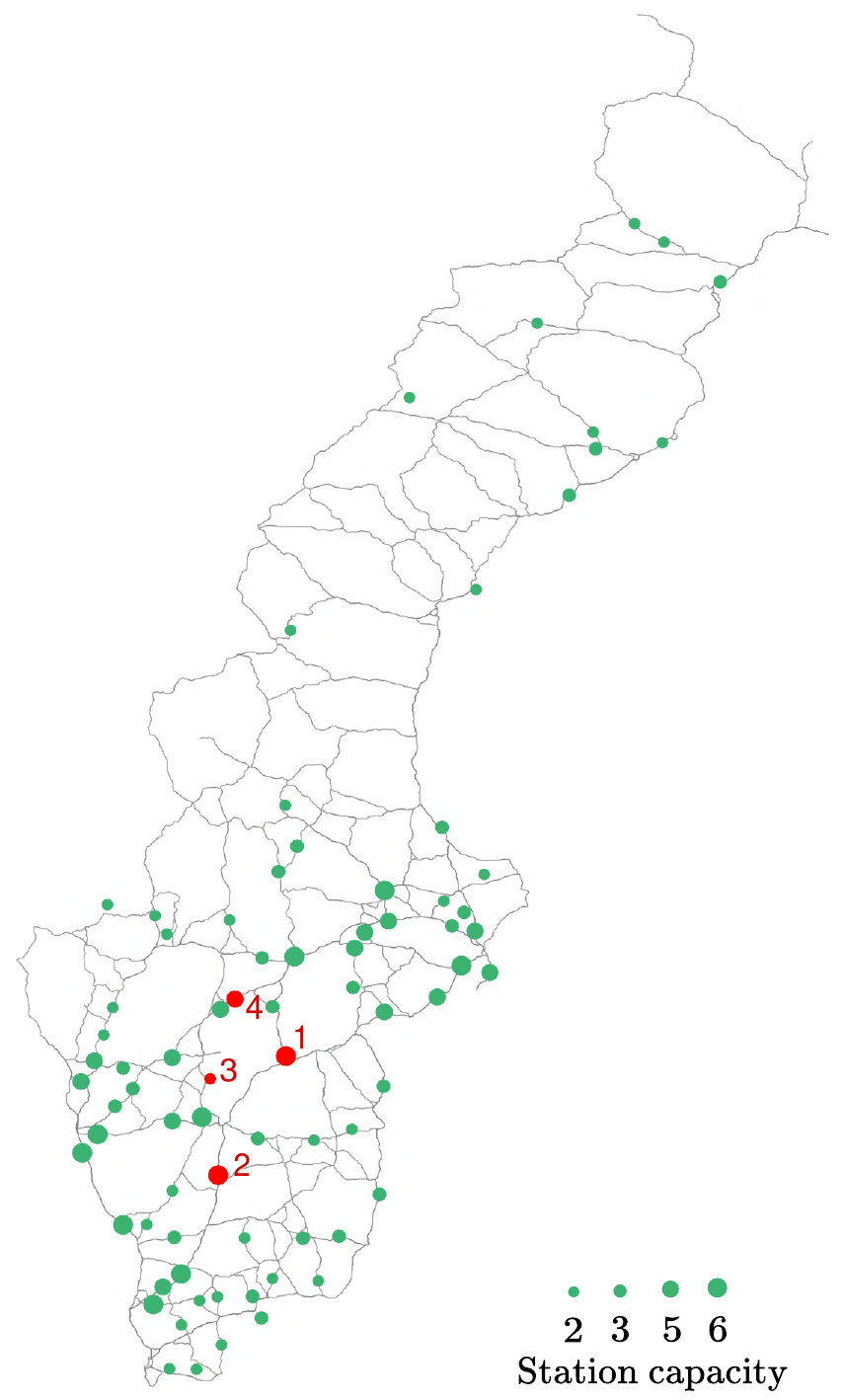}}
\subfigure[Route of a single truck]
{\includegraphics[width=0.32\textwidth]{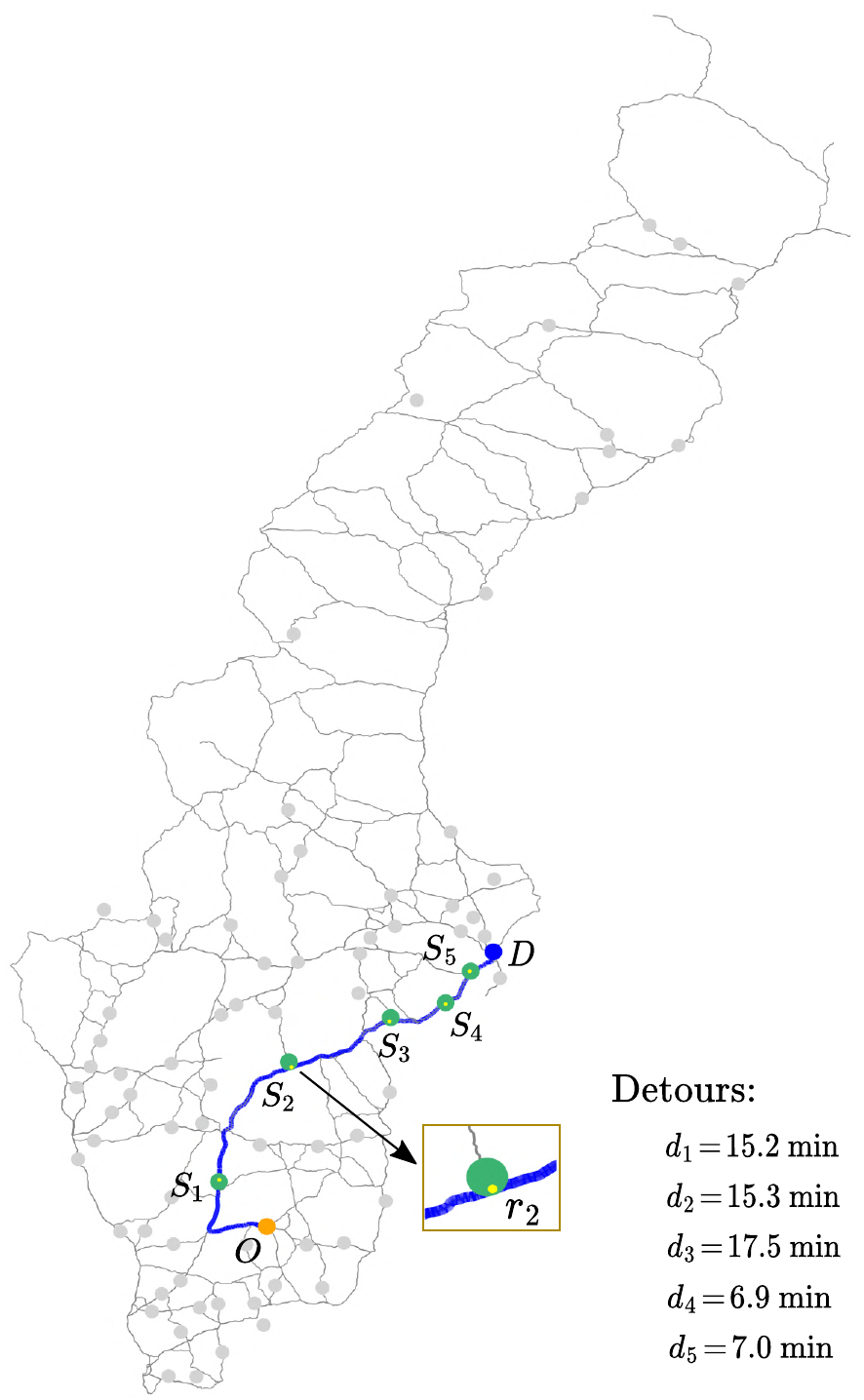}}
\end{minipage}
\vspace{-8pt}
\caption{(a) The transport flow among different regions in Sweden. (b) The potential charging stations for electric trucks, where the size of the green nodes denotes the number of charging ports (i.e., capacity) at the station. Four representative charging stations are marked in red and their waiting time plots are shown in Figure~\ref{Fig.6}. (c) The route model of a single truck.}
\label{Fig.5}
\end{figure*}

\begin{figure*}[t]
     \centering
     \includegraphics[width=1.0\linewidth]{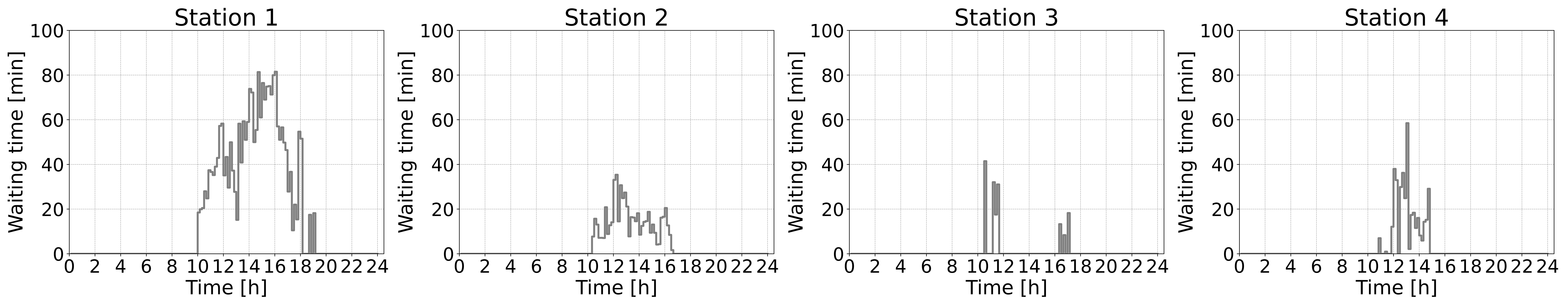}
      \vspace{-15pt}
      \caption{The waiting time forecast models at four representative charging stations that are marked in red in Figure~\ref{Fig.5}(b). The models are obtained using the historical data at each local station.}
      \label{Fig.6}
\end{figure*}

This section tests the proposed coordination framework over the Swedish road network using realistic road and truck data. We first introduce the simulation setup, including practical transport flows between different regions, realistic delivery routes, and in-use parameter settings for electric trucks. In the sequel, we provide the evaluation and comparison results of the simulation studies that illustrate the effectiveness of the developed charging coordination approach. The code for some sample implementations is available online.\footnote{See {https://github.com/kth-tingbai/Charging-Coordination-under-Limited-Facilities}.}

\subsection{Setup}
\subsubsection{Missions, Routes, and Charging Stations} We consider the Swedish road network comprising $105$ real road terminals, as shown by the blue nodes in Figure~\ref{Fig.5}(a). The transport flow between any two nodes represents the delivery demands between two corresponding regions, which is derived from the SAMGODS model and built upon realistic producer and consumer data in Sweden. We then generate the delivery missions for $1,000$ trucks, where the origin-destination (OD) pair of each truck is chosen from the set of road terminals. To achieve a realistic distribution of the delivery missions, the transport flow given in Figure~\ref{Fig.5}(a) is employed. Specifically, the probability of selecting two terminals $i$ and $j$ as an OD pair is computed by $F_{i,j}/\sum_{i,j}F_{i,j}$, where $F_{i,j}$ denotes the truck flow from the terminal $i$ to $j$ based on the SAMGODS model. Given the realistic coordinates of the nodes comprising the OD pairs, the route for each truck can be planned using the open-source routing service \textit{OpenStreetMap}~\cite{OpenStreetMap}.

As there are very few charging stations in use nowadays for commercial electric trucks, we make use of some extra road terminals obtained from the SAMGODS model as potential charging stations in the network, as shown by the green nodes in Figure~\ref{Fig.5}(b). Using a certain search range, one can identify a collection of charging stations along the pre-planned route for each trip, as well as the ramps in the route leading to the shortest detour to reach the corresponding charging stations. The capacity of a charging station is determined so that it is proportional to the number of trucks that may use the station for charging. By ``may use,” we mean that the truck has the station along its pre-planned route. More details on the size of each station are provided in Figure~\ref{Fig.5}(b). Furthermore, trucks' nominal travel times on its route segments $\{\tau_k\}_{k=0}^N$ and for detours $\{d_k\}_{k=1}^N$ are obtained from the \textit{OpenStreetMap}. As an illustration, the route model of one truck is shown in Figure~\ref{Fig.5}(c).

\subsubsection{Parameter Settings} We model the departure times of trucks to be random within the time window 07:00–10:00, stimulating the typical schedule of freight transportation operations. In addition, we assume each truck commences its journey with an arbitrary but feasible battery from its origin. Parameters for the electric trucks are set using the latest data published by Scania~\cite{ElectricTruck}. More specifically, we consider electric trucks with a maximum capacity of $40$ tonnes and equipped with an installed battery of $624$~kWh, among which $468$~kWh is usable. With this, trucks can cover a travel distance of $350$ km on a single charge. The maximum acceptable charging power of a truck is $350$~kW, while the charging power provided by each station is $300$~kW. For safe operation, $e_s$ is set as $25\%$ of the installed battery capacity. In addition, we assume a constant driving speed of $82$~km/h for every truck, leading to a battery consumption rate of approximately $1.83$~kWh/min. Charging at each station incurs the electricity cost of $0.36$\texteuro/kWh, and the monetary labor compensation due to extended travel time is considered as $2$\texteuro/min. The penalty for the violation of the delivery deadline is set as $10$\texteuro/min.
\begin{figure}[t]
     \centering
     \includegraphics[width=0.965\linewidth]{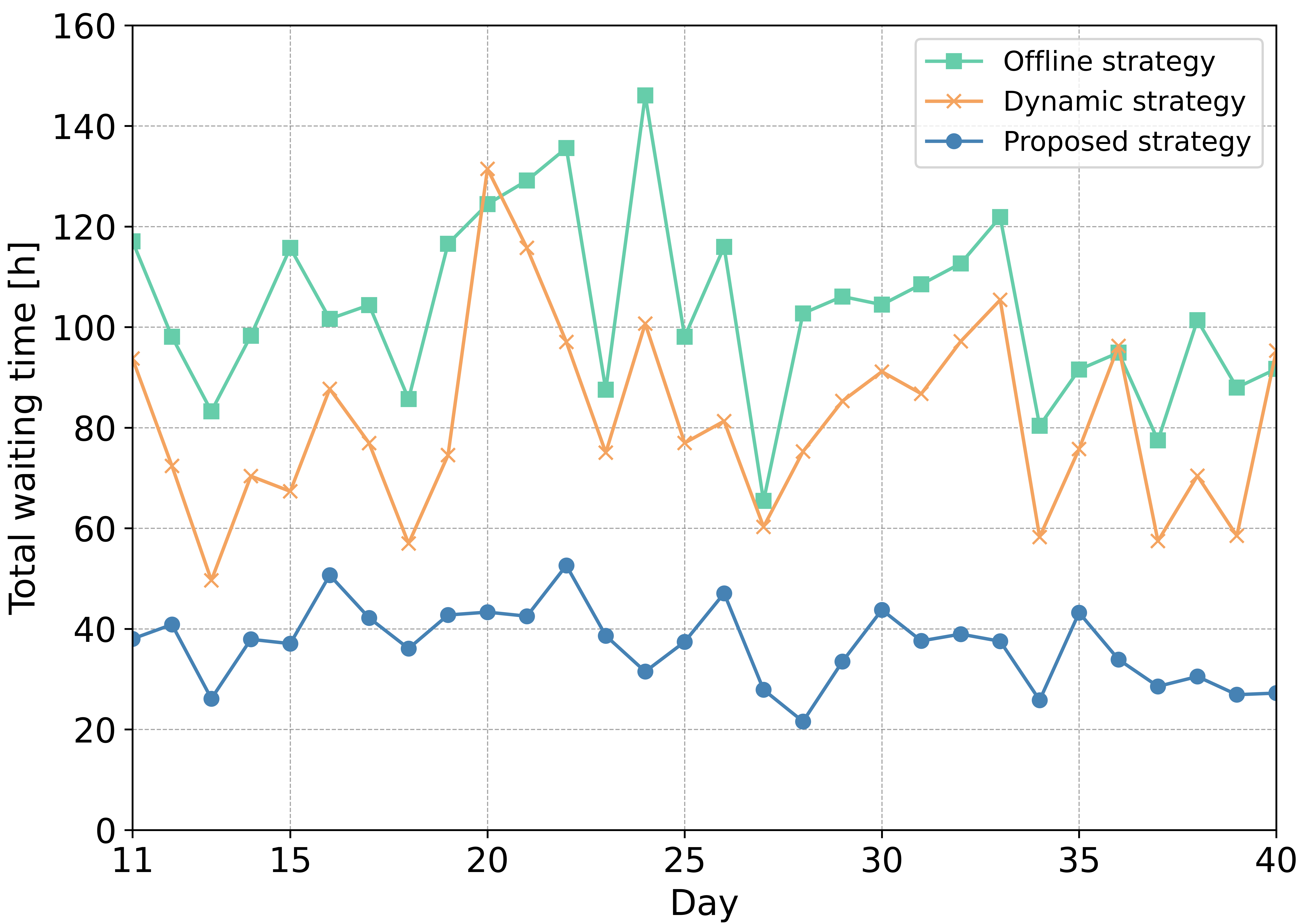}
      \caption{The total waiting time in the deterministic scenario.}
      \label{Fig.7}
\end{figure}

\subsection{Results and Comparison}
To test the developed coordination framework, we compare our method with two other coordination approaches: the \emph{offline strategy} and the \textit{dynamic strategy}. The offline strategy is computed by each truck offline without considering charging congestion at stations or travel uncertainties. The dynamic strategy, similar to our proposed coordination method, is computed by each truck whenever it arrives at a ramp. However, in the dynamic strategy, communication occurs only between the truck and its nearby station, assuming zero waiting times at distant stations. Moreover, we compare the deterministic and stochastic scenarios in our simulation study. In the deterministic scenario, all parameters and related variables are certain. In the stochastic scenario, we assume that trucks' travel times and battery consumption are subject to $3\%$, $5\%$, and $7\%$ disturbances. In the subsequent comparisons, we focus on $\delta_k^1\!=\!\delta_k^2\!=\!5\%$ for all $k$. The results for the other two cases are similar and thus omitted. Note that while travel uncertainties may vary across different route segments, our method can readily accommodate this by adjusting the values of $\delta_k^1$ and $\delta_k^2$ with $k$ to align with real-world conditions. Below, we present the simulation results in each scenario.

\begin{figure}[t]
     \centering
     \includegraphics[width=1.0\linewidth]{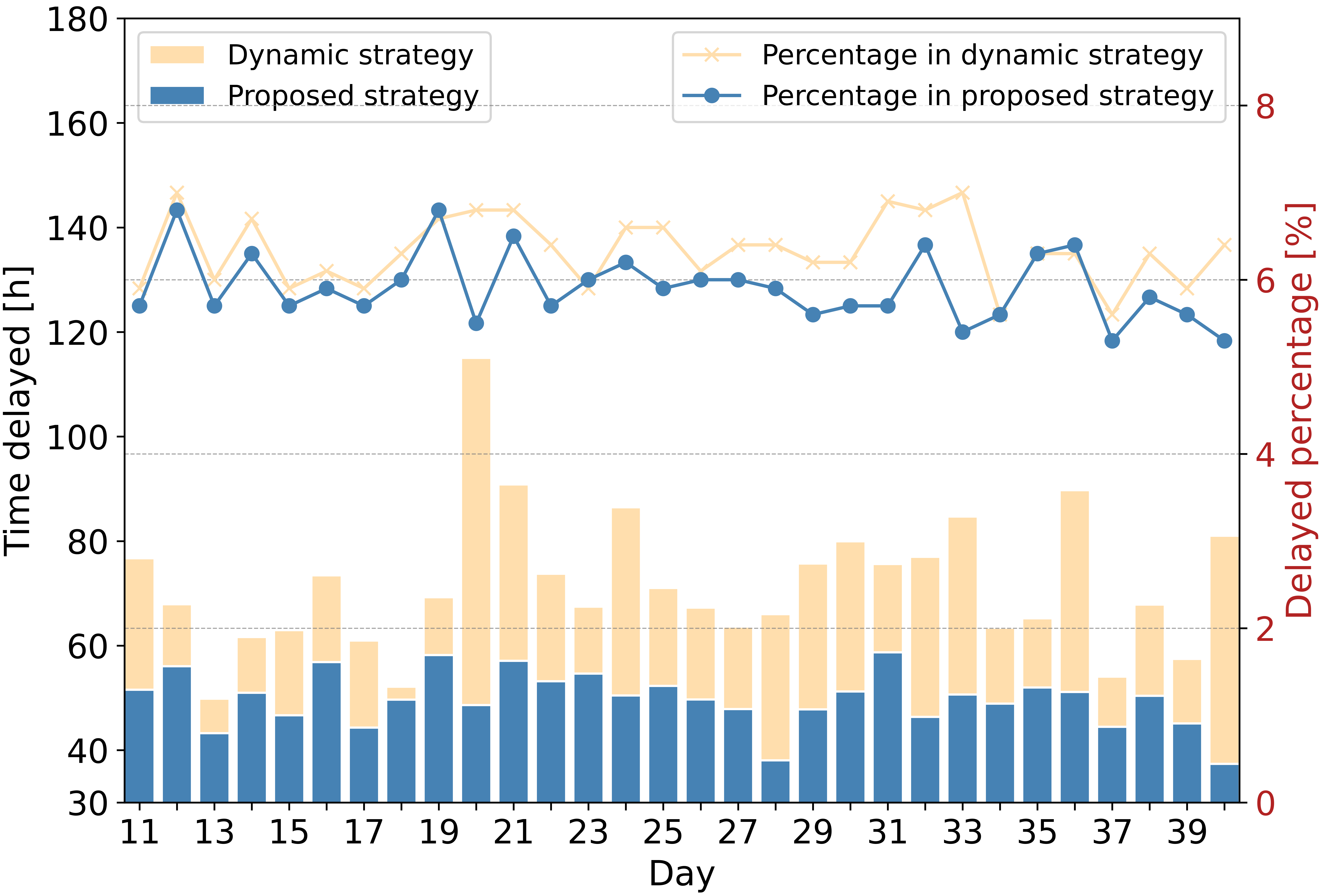}
      \caption{The cumulative time delay in the deterministic scenario.}
      \label{Fig.8}
\end{figure}

\subsubsection{Deterministic Scenario}
We conducted simulations for $1,000$ trucks traversing the Swedish road network over $40$ days. During the initial $10$ days, the stations collect arrival and waiting time data for trucks, setting the estimated waiting times $\Tilde{w}$ as zero for all distant trucks. Using the data collected during this phase, the stations construct waiting time forecast models as waiting time estimation plots. For example, the waiting time forecast models of the four charging stations marked in red in Figure~\ref{Fig.5}(b) are shown in Figure~\ref{Fig.6}. In each waiting time estimation plot, the timeline per day is discretized into $5$ minute intervals when recording trucks' arrivals and actual waiting times. The y-axis in Figure~\ref{Fig.6} shows the average waiting time of the trucks that arrived at the station during corresponding time periods. The obtained waiting time forecast models are then used by the stations to compute the estimated waiting times for trucks upon their requests when applying the proposed method in the following $30$ days. 

Simulation results for the deterministic scenario are shown in Figure~\ref{Fig.7} and Figure~\ref{Fig.8}. Specifically, Figure~\ref{Fig.7} presents the total waiting times of $1,000$ trucks each day from day $11$ to day $40$, compared among the three charging strategies. As shown in the figure, the dynamic charging strategy outperforms the offline strategy in most cases, attributed to its dynamic updates utilizing accurate waiting time information provided by the nearby station. It may also happen that the dynamic strategy performs worse than the offline strategy, see, for instance, on day $20$. This indicates that accurate waiting time estimations at distant stations are crucial for optimizing the charging strategy. In contrast to the dynamic strategy, our method enables more precise estimations of waiting times at distant stations by providing estimated arrival-time windows to distant stations. This leads to substantial reductions in trucks' total waiting times, as shown in Figure~\ref{Fig.7}. 

Figure~\ref{Fig.8} shows the cumulative delay experienced by all trucks relative to their delivery deadlines, compared between the dynamic and proposed charging strategies. While the results vary case-by-case, our method generally results in less overall delay. Additionally, the simulation result indicates that approximately $6\%$ of the trucks experience delays in fulfilling their transportation tasks under the two charging strategies.

\subsubsection{Stochastic Scenario}
\begin{figure}[t]
     \centering
     \includegraphics[width=0.965\linewidth]{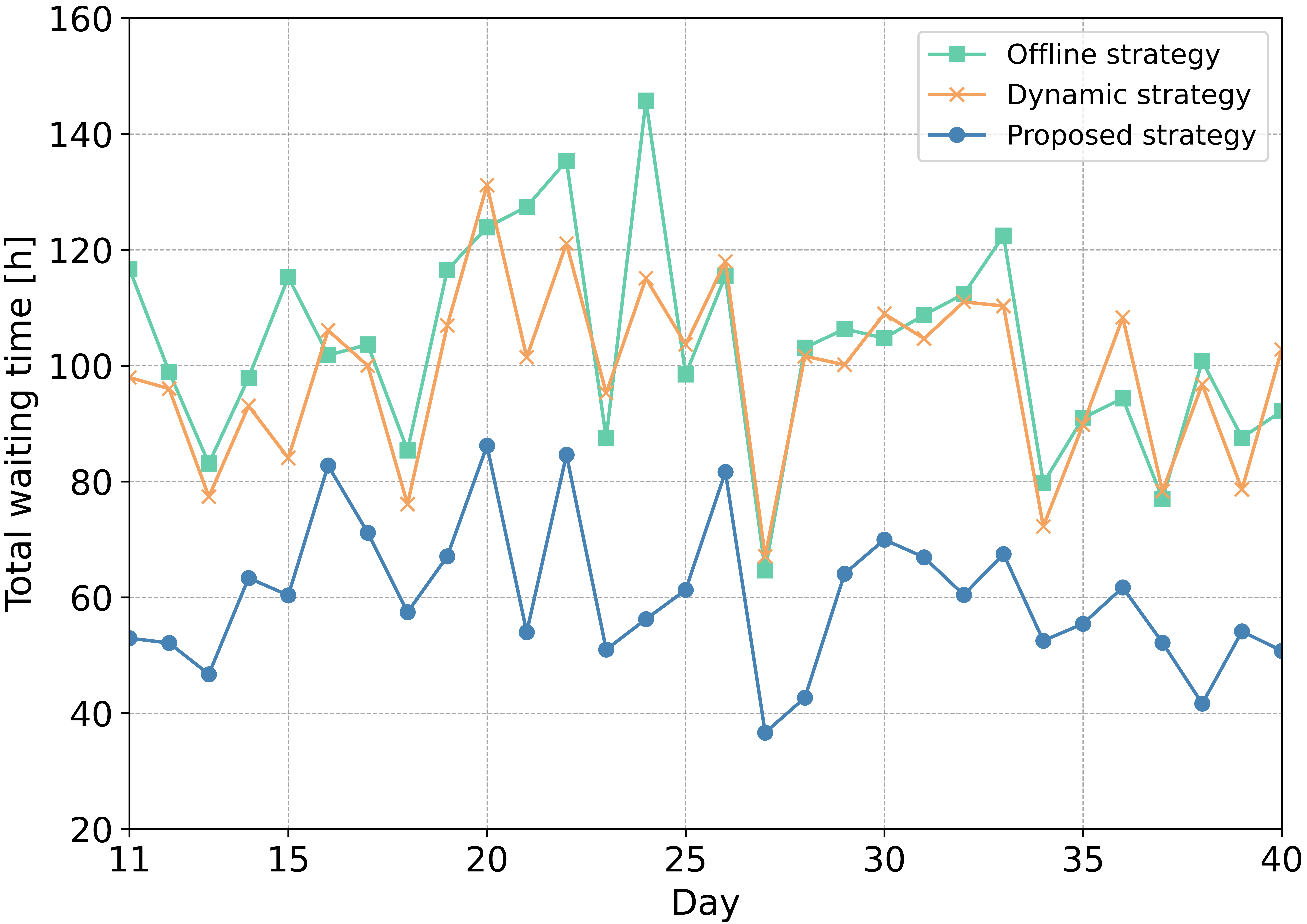}
     \vspace{-5pt}
      \caption{The total waiting time in the stochastic scenario.}
      \label{Fig.9}
\end{figure}

\begin{figure}[t]
     \centering
     \includegraphics[width=1.0\linewidth]{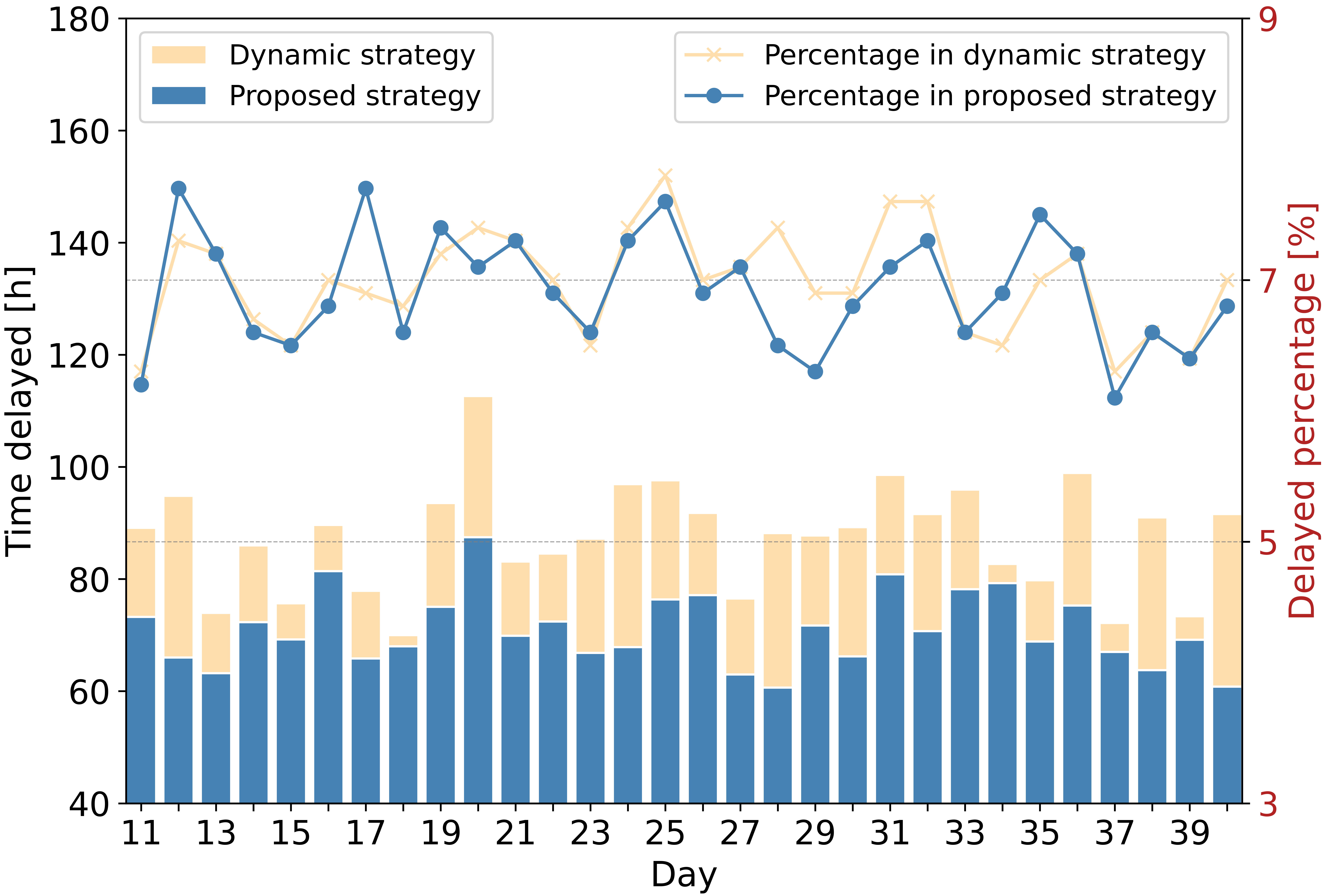}
     \vspace{-16pt}
      \caption{The cumulative time delay in the stochastic scenario.}
      \label{Fig.10}
\end{figure}

\begin{figure}[t]
     \centering
     \includegraphics[width=0.963\linewidth]{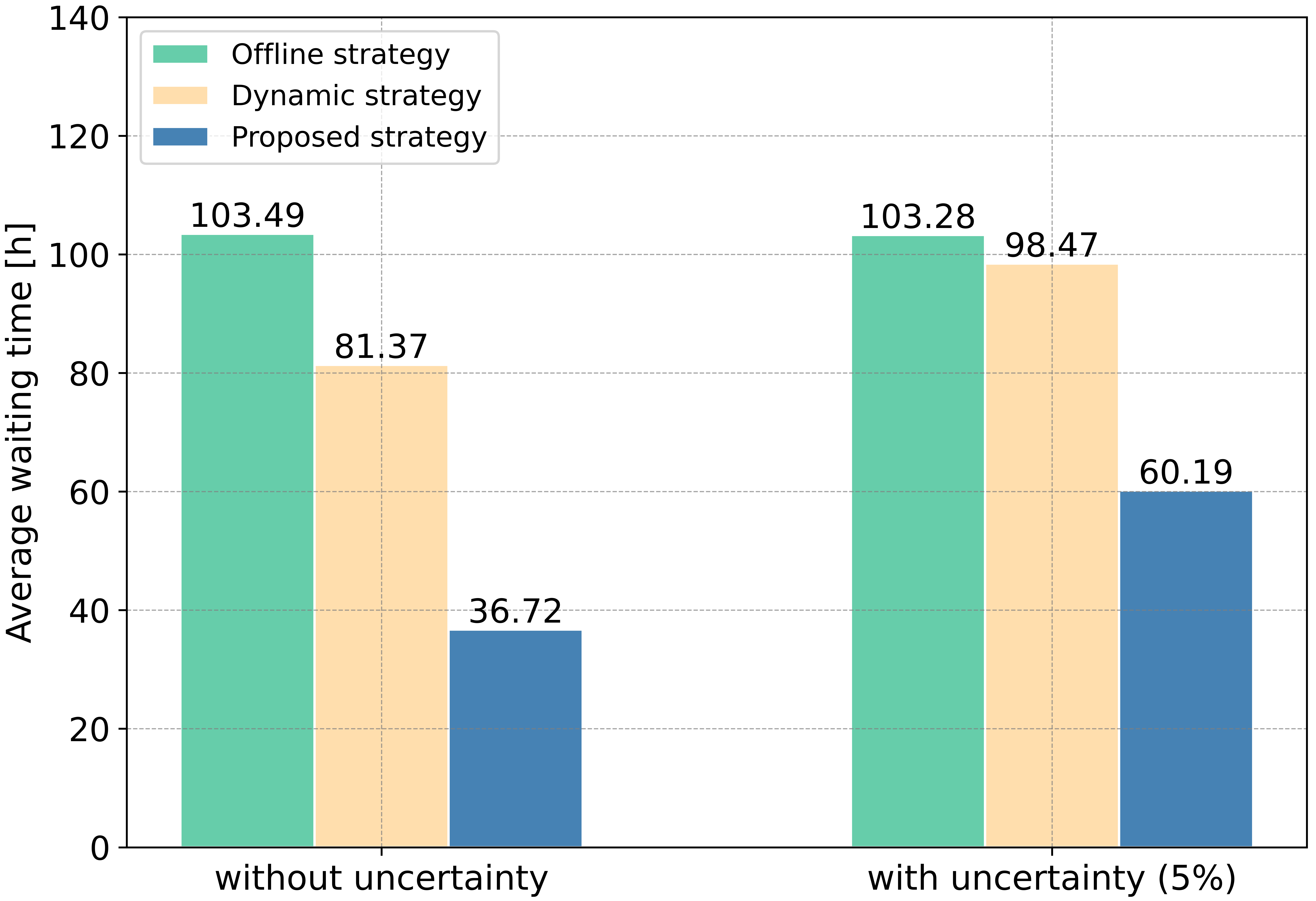}
      \caption{Comparison of the average waiting time of all trucks per day.}
      \label{Fig.11}
\end{figure}

\begin{figure}[t]
     \centering
     \includegraphics[width=0.94\linewidth]{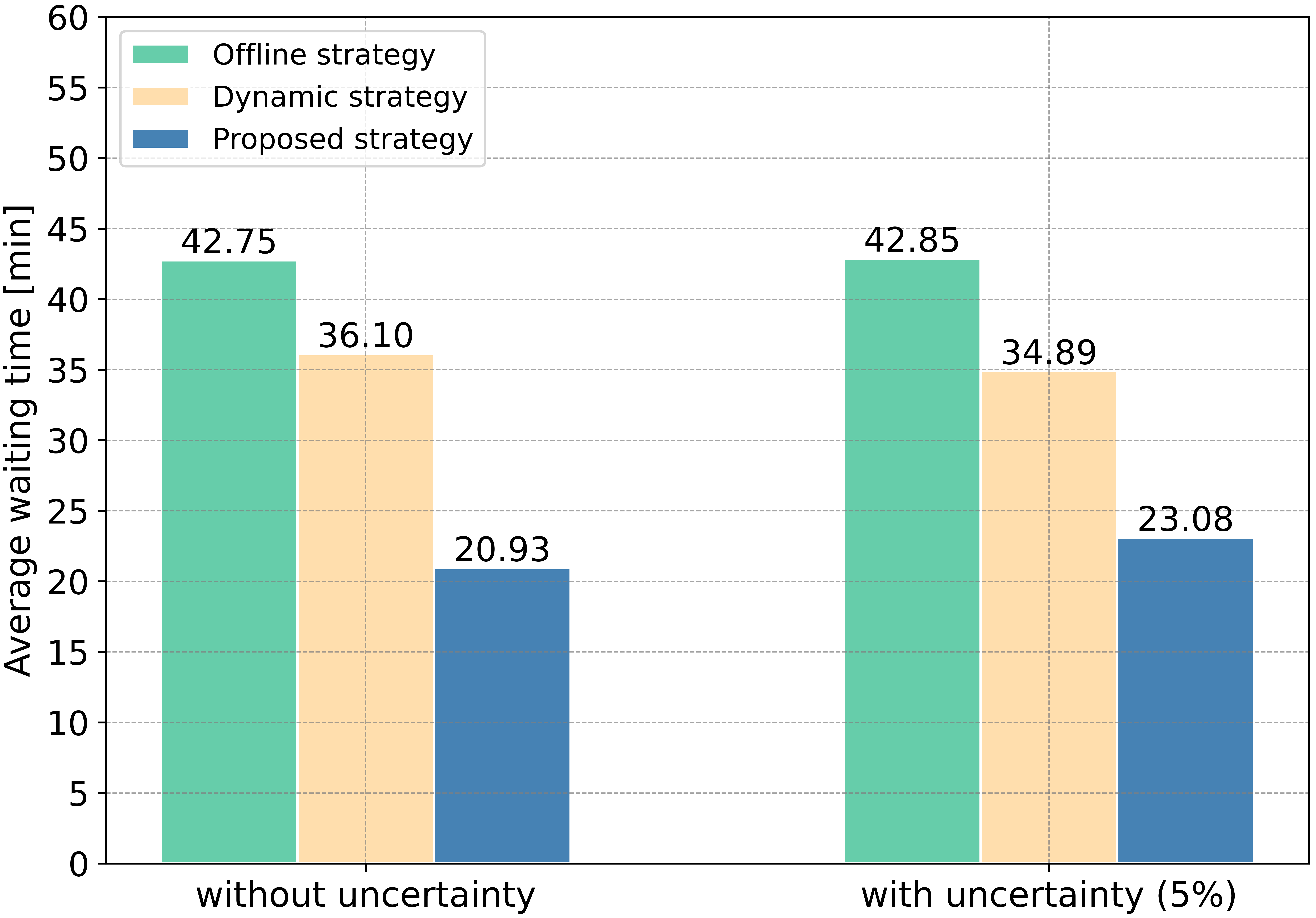}
      \caption{Comparison of the average waiting time per truck per day.}
      \label{Fig.12}
\end{figure}

To further evaluate the resilience of the developed coordination framework, we performed simulation studies under a stochastic scenario, considering $5\%$ uncertainties in travel times and battery consumption, i.e., setting $\delta_k^1\!=\!\delta_k^2\!=\!5\%$ for all $k$. Apart from these uncertainties, the other parameter settings remain identical to those of the deterministic scenario. In this case, the framework introduced in our previous work \cite{bai2024distributed} does not guarantee feasible charging plans for the trucks. The dynamic strategy we compare against can be viewed as an enhanced version of the framework in~\cite{bai2024distributed}, as it guarantees plan feasibility by using the charging planning method presented in Section~\ref{Subsection IV.C}.

As illustrated in Figure~\ref{Fig.9}, the dynamic charging strategy yields little improvement over the offline strategy when incorporating travel uncertainties. In addition, we observe that the offline strategy achieves highly similar performance in both deterministic and stochastic scenarios. This consistency could be due to the offline charging plan being developed independently of the environmental impacts, thereby exhibiting strong robustness. The results also show that, with bounded travel uncertainties, the proposed method effectively reduces waiting times caused by charging congestion. 

In the stochastic scenario, the total travel time delay for all trucks in the network increases in both the dynamic and our proposed strategies, as shown in Figure~\ref{Fig.10}. The proportion of delayed trucks fluctuates across different cases, hovering around $7\%$, and the amplitude of the fluctuation turns bigger than that of the deterministic scenario. 

\subsubsection{Overall Comparison}
Towards providing an overall comparison and analysis, we show in Figure~\ref{Fig.11} and Figure~\ref{Fig.12} the average waiting time of trucks in different charging strategies, compared between deterministic (i.e., without uncertainty) and stochastic (i.e., with uncertainty) scenarios. More specifically, Figure~\ref{Fig.11} shows the average daily waiting time for all trucks, while Figure~\ref{Fig.12} illustrates the average daily waiting time per truck, focusing only on those with non-zero waiting times. 

As shown in Figure~\ref{Fig.11}, the proposed charging coordination approach reduces averaged waiting times for all trucks per day by approximately $64.5\%$ (i.e., $(103.49-36.71)\!\times\!{100\%}/103.49$) in the deterministic scenario and $41.7\%$ in the stochastic scenario compared to the offline strategy. In contrast, the dynamic strategy achieves reductions of about $21.4\%$ and $4.7\%$, respectively. For individual trucks, Figure~\ref{Fig.12} indicates that our method reduces the average waiting time per truck by approximately $51\%$ in deterministic scenarios and $46.1\%$ in scenarios with $5\%$ travel uncertainties in comparison with the offline strategy. While the dynamic strategy results in an $18.6\%$ reduction in waiting time in the stochastic scenario, our proposed strategy enhances this improvement by an additional $33.8\%$ (i.e., $(34.89\!-\!23.08)\!\times\!{100\%}/34.89$).

\subsubsection{Operational Cost} To evaluate the economic benefits of the proposed approach, Figure~13 presents the daily average operational costs for $1,000$ trucks, comparing the dynamic and proposed coordination methods under varying levels of uncertainty. The results demonstrate that, while the average operational costs for both methods increase with higher uncertainty, our approach consistently achieves lower total costs than the dynamic method. Specifically, the proposed method reduces the average operational costs by approximately $10.4\%$ in the deterministic scenario and by about $6.5\%$ at an uncertainty level of $7\%$. Moreover, the percentage of delayed trucks is consistently lower with our method, ranging from $5.9\%$ to $7\%$ as uncertainty increases, compared to the higher delay rates observed with the dynamic approach.
\begin{figure}[t]
     \centering
     \includegraphics[width=1.0\linewidth]{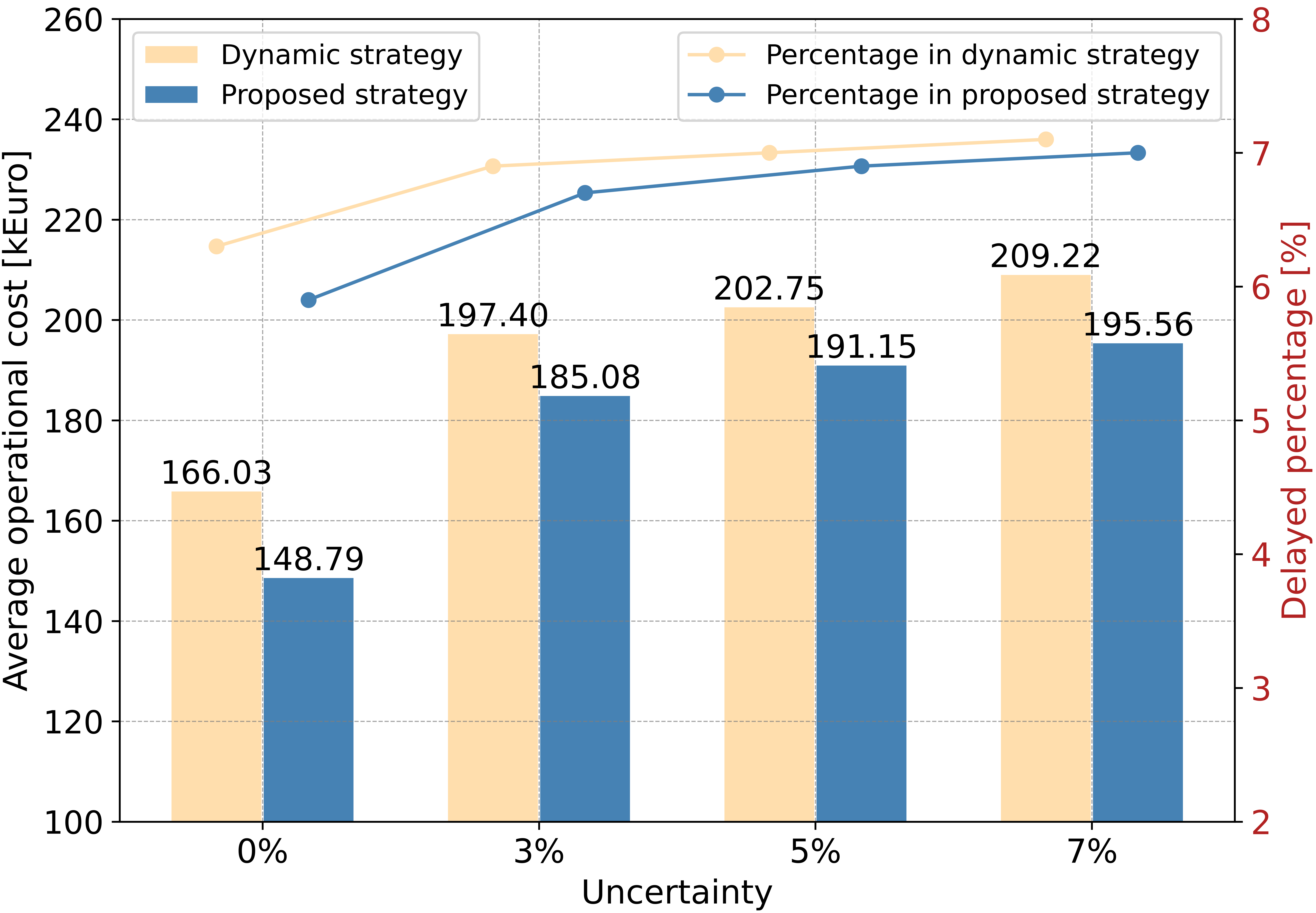}
     \vspace{-16pt}
      \caption{The average operational cost of all trucks per day.}
      \label{Fig.13}
\end{figure}

\subsubsection{Evaluation to Stations} 
\begin{figure}[!t]
    \centering
    \subfigure[Without uncertainty]{
    \includegraphics[width=8.435cm,height=4.0cm]{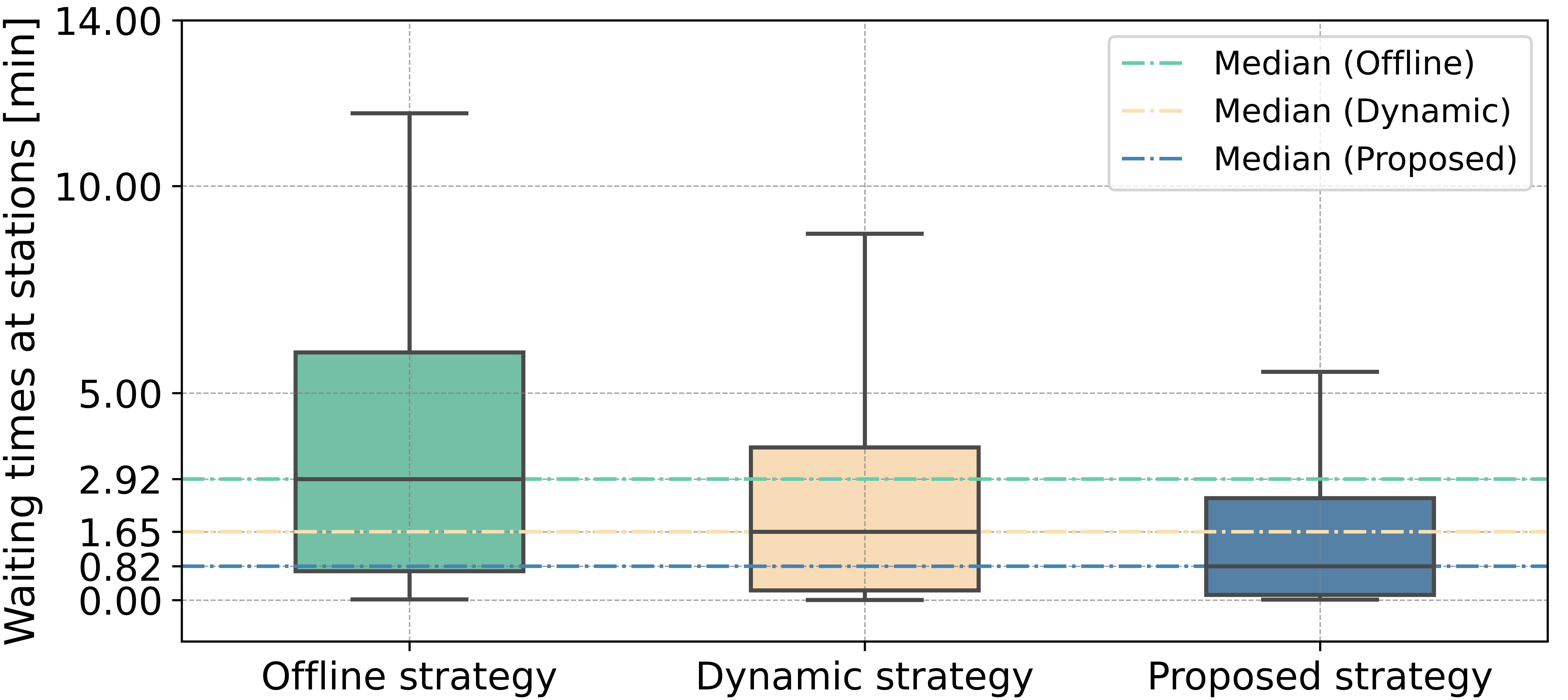}
    }
    \centering
    \subfigure[With uncertainty ($5\%$)]{
    \includegraphics[width=8.435cm,height=4.0cm]{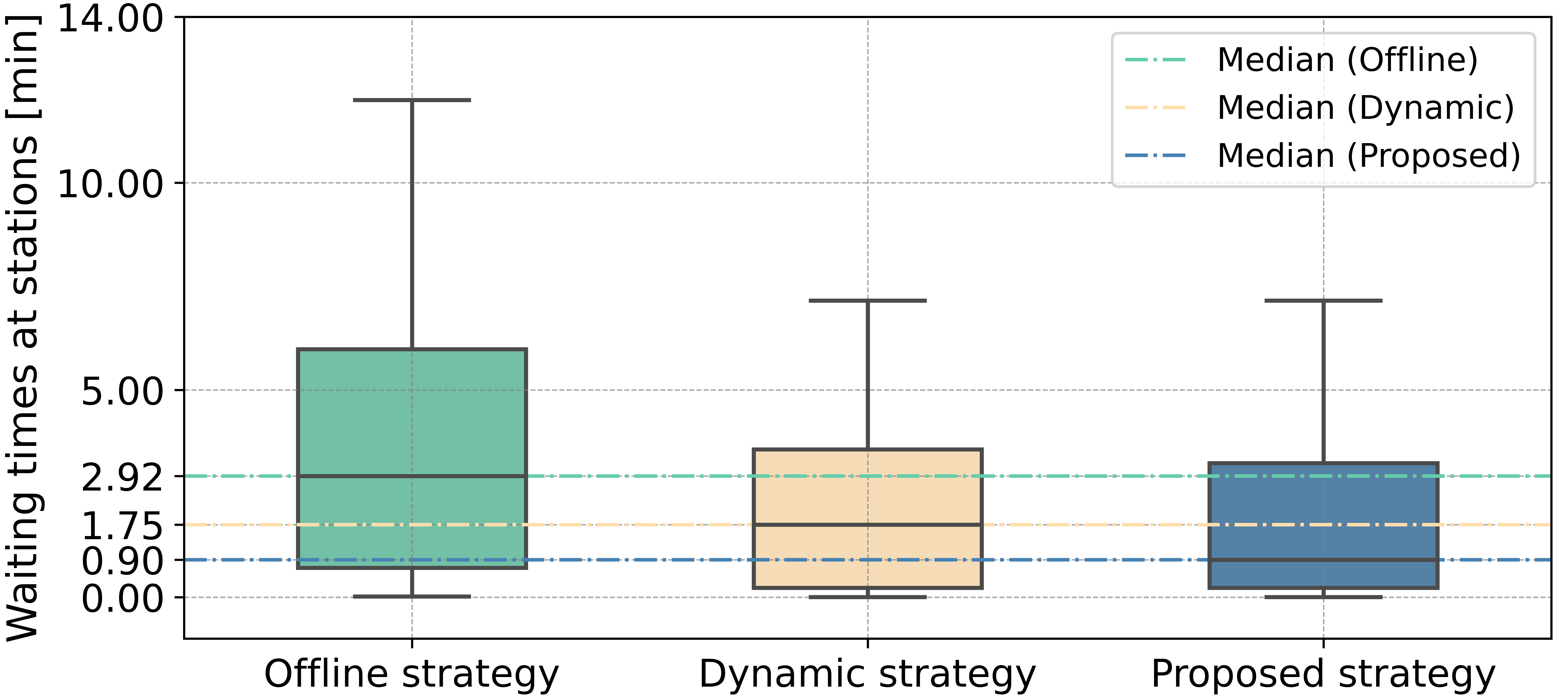}
    }
    \DeclareGraphicsExtensions.
    \caption{Trucks' average waiting times at stations, compared among different charging coordination strategies (a) without and (b) with travel uncertainty.}
    \label{Fig.14}
\end{figure}

The charging congestion at each station is then evaluated by computing the average waiting time per truck at the station. Figure~\ref{Fig.14}(a) illustrates trucks' average waiting times at every station in the deterministic scenario. The medians of the average waiting times per truck in the offline, dynamic, and proposed methods are $2.92$, $1.65$, and $0.82$ minutes, respectively. The corresponding interquartile ranges are $5.28$, $3.45$, $2.34$, indicating the range of the middle $50\%$ of the data. For presentation clarity, outliers have been removed from the figure. These results demonstrate that the proposed coordination method reduces charging congestion significantly compared to the other two methods, thereby enhancing charging efficiency at the stations. 

The results for the stochastic scenario are presented in Figure~\ref{Fig.14}(b). As is shown, the median average waiting times for the dynamic and proposed strategies with uncertainty are slightly higher than those without uncertainty, while the offline strategy shows greater robustness to uncertainty. In addition, the interquartile range and whiskers for the proposed strategy are wider in the stochastic scenario, indicating a larger distribution range due to travel uncertainties. Nevertheless, our coordination strategy still outperforms both the offline and dynamic strategies in alleviating charging congestion. 

\section{Conclusions}\label{Section VI}
We have addressed the problem of charging coordination between electric trucks and charging stations. Our goal is to minimize individual trucks' operational costs while reducing their waiting times at the stations. To this end, we have designed a distributed coordination framework that supports information exchange between the trucks and the stations, and enables trucks to make their own charging plans dynamically upon approaching stations. This framework extends our previous work by introducing three key components. The first two components are waiting time forecast models computed by the stations, and the estimated arrival-time windows computed by the trucks. These two components contribute to effective estimations of waiting times, which are used by the trucks for charging planning. The third component is a new charging planning approach by the trucks that handles bounded travel uncertainties. To test the effectiveness of our proposed scheme, we have conducted simulation studies involving $1,000$ trucks traversing the Swedish road network for $40$ days. When there are $5\%$ uncertainties in the traveling time and energy consumption, our proposed scheme reduces the average waiting time per truck by approximately $46.1\%$ when compared with offline charging planning without coordination, and by about $33.8\%$ when compared with the coordination that does not involve forecast model and arrival-time windows computations. 

In contrast to other related works handling uncertainties, our approach does not assume a probability distribution for stochastic uncertainties, is less conservative than robust optimization, and is computationally efficient, leveraging the boundedness of uncertainty quantities. As for future works, we would like to incorporate nonlinear battery dynamics into the coordination framework and investigate the potential of machine learning techniques in computing the forecast models.

\section*{Acknowledgment}
The authors would like to express their sincere gratitude to Prof. Dimitri P. Bertsekas for his valuable comments and suggestions. We would also like to thank Albin Engholm for providing simulation data from the SAMGODS model.

\appendix
\begin{proofproposition1} 
Let $(b_k^*,t_k^*)$ be the optimal charging decision taken by the truck at station $S_k$. In addition, let $\{(\hat{b}_h^k,\hat{t}_h^k)\}_{h=k}^{\ell-1}$, $\ell\!=\!k\!+\!1,\dots,N$, be the optimal solution of the problem \eqref{Eq.14} and \eqref{Eq.15} computed at ramp $r_k$, and let $\{(\hat{b}_h^{k+1},\hat{t}_h^{k+1})\}_{h=k+1}^{\ell-1}$, $\ell\!=\!k\!+\!2,\dots,N$, be the one computed at ramp $r_{k+1}$. The effective charging power at $S_{k+1}$ is denoted as $P_{k+1}^c\!=\!\min\{P_{k+1},P_{\max}\}$. We start with the proof of \eqref{Eq.20a}.
\begin{itemize}
\item [(a)] The proof consists of two steps. First, we will prove
\begin{align}
\!\!\!\!\Big\{\!(b_k^*,t_k^*), \!(\hat{b}_{k+1}^{k+1},\hat{t}_{k+1}^{k+1}\!\!-\!\Delta{\hat{t}}_{k+1}),\!\{\!(\hat{b}_{h}^{k+1}\!,\hat{t}_h^{k+1})\!\}_{h=k+2}^{\ell-1}\!\Big\}\label{Eq.26}
\end{align}
is a feasible solution to the problem \eqref{Eq.14} and \eqref{Eq.15} computed at ramp $r_k$, where 
\begin{align}
\Delta{\hat{t}}_{k+1}\!=\!\frac{(\delta_k^2\bar{P}\tau_k-\delta_{e,k})}{P_{k+1}^c}\geq{0}\label{Eq.27}
\end{align}
with $\delta_{e,k}\!\in\![-\delta_k^2\bar{P}\tau_k,\delta_k^2\bar{P}\tau_k]$. Without loss of generality, we assume that $\hat{b}_{k+1}^{k+1}\!=\!1$ and $\hat{t}_{k+1}^{k+1}\!\geq\!\Delta{\hat{t}}_{k+1}$. The case for $\hat{b}_{k+1}^{k+1}\!=\!0$ will be discussed later. Since $(b_k^*,t_k^*)$ is optimal to the problem \eqref{Eq.21}-\eqref{Eq.23}, it meets the constraints in~\eqref{Eq.15}. To proceed, we denote by $e_{k+1}^*$ the remaining battery of the truck at ramp $r_{k+1}$ computed by the model \eqref{Eq.15b} applying the optimal control $(b_k^*,t_k^*)$ and denote as $\hat{e}_{k+1}$ its actual remaining battery. By \eqref{Eq.15b}, we have
\begin{align}
\hat{e}_{k+1}=e_{k+1}^*-\delta_k^2\bar{P}\tau_k+\delta_{e,k}.\label{Eq.28}
\end{align}
Since $(\hat{b}_{k+1}^{k+1},\hat{t}_{k+1}^{k+1})$ is computed at $r_{k+1}$ with the battery $\hat{e}_{k+1}$, constraints \eqref{Eq.15d} and \eqref{Eq.15e} are satisfied, i.e., 
\begin{align}
\hat{t}_{k+1}^{k+1}P_{k+1}^c\leq{e_f\!-\!(\hat{e}_{k+1}\!-\!\bar{P}d_{k+1})}.\label{Eq.29}
\end{align}
In light of \eqref{Eq.27}-\eqref{Eq.29}, we have
\begin{align*}
e_f\!-\!(e_{k+1}^*\!\!-\!\bar{P}d_{k+1})\!&=\!e_f\!-\!(\hat{e}_{k+1}\!\!-\!\bar{P}d_{k+1})\!-\!(\delta_k^2\bar{P}\tau_k\!\!-\!\delta_{e,k})\\
&\geq{\big(\hat{t}_{k+1}^{k+1}\!-\!\Delta{\hat{t}}_{k+1}\big)P_{k+1}^c}.
\end{align*}
Thereby, \eqref{Eq.15d} and \eqref{Eq.15e} hold for $(\hat{b}_{k+1}^{k+1},\hat{t}_{k+1}^{k+1}\!-\!\Delta{\hat{t}}_{k+1})$ and $e_{k+1}^*$. Below, we will prove that the remaining battery at ramp $r_{k+2}$ computed at ramp $r_k$, denoted by $e_{k+2}^*$, is the same as that computed at ramp $r_{k+1}$, denoted as $\hat{e}_{k+2}$. By \eqref{Eq.15b}, one has
\begin{subequations}
\label{Eq.30}
\begin{align}
e_{k+2}^*&=e_{k+1}^*\!+\hat{b}_{k+1}^{k+1}\big(\hat{t}_{k+1}^{k+1}\!-\!\Delta{\hat{t}}_{k+1}\big)P_{k+1}^c\!+\!\delta_{k+1}^2\bar{P}\tau_{k+1}\nonumber\\
&\quad-\!\bar{P}\big(2\hat{b}_{k+1}^{k+1}d_{k+1}\!+\!\tau_{k+1}\big),\label{Eq.30a}\\
\hat{e}_{k+2}&=\hat{e}_{k+1}\!+\hat{b}_{k+1}^{k+1}\hat{t}_{k+1}^{k+1}P_{k+1}^c\!+\delta_{k+1}^2\bar{P}\tau_{k+1}\nonumber\\
&\quad-\!\bar{P}\big(2\hat{b}_{k+1}^{k+1}d_{k+1}\!+\!\tau_{k+1}\big).\label{Eq.30b}
\end{align}
\end{subequations}
According to \eqref{Eq.27}, \eqref{Eq.28} and \eqref{Eq.30}, it derives
\begin{align*}
e_{k+2}^*-\hat{e}_{k+2}=(\delta_k^2\bar{P}\tau_k\!-\!\delta_{e,k})(1\!-\hat{b}_{k+1}^{k+1})={0}.
\end{align*}
Thus, $e_{k+2}^*\!=\!\hat{e}_{k+2}$ holds and \eqref{Eq.15c} is satisfied by $e_{k+2}^*$. This proves the feasibility of $\{\hat{b}_{h}^{k+1},\hat{t}_h^{k+1}\}_{h=k+2}^{\ell-1}$. Given the above, we prove that \eqref{Eq.26} is a feasible solution to the problem \eqref{Eq.14} and \eqref{Eq.15} solved at ramp $r_k$. 

We are now ready to prove \eqref{Eq.20a}. By definition, $\{(\hat{b}_h^k,\hat{t}_h^k)\}_{h=k}^{\ell-1}$ is optimal to \eqref{Eq.14} and achieves the minimum cost, i.e.,
\begin{align}
&\sum_{h=k}^{\ell-1}\hat{b}_h^k\big(2d_h\!+\hat{t}_h^k\big)\nonumber\\
&\leq{b_k^*\big(2d_k\!+\!t_k^*\big)+\!\!\!\sum_{h=k+1}^{\ell-1}\!\!\hat{b}_h^{k+1}\big(2d_h\!+\!\hat{t}_h^{k+1}\big)}\!-\!\hat{b}_{k+1}^{k+1}\Delta{\hat{t}}_{k+1}\nonumber\\
&\leq{b_k^*\big(2d_k\!+\!t_k^*\big)+\!\!\!\sum_{h=k+1}^{\ell-1}\!\!\hat{b}_h^{k+1}\big(2d_h\!+\!\hat{t}_h^{k+1}\big)}.\label{Eq.31}
\end{align}
By \eqref{Eq.16}, for $\ell\!=\!k\!+\!2,\dots,N$, we have
\begin{subequations}
\label{Eq.32}
\begin{align}
\underline{a}_{\ell}^{k+1}\!\!&=\!a_{k+1}\!+\!\!\!\!\sum_{h=k+1}^{\ell-1}\!\!\!\!\Big(\hat{b}_h^{k+1}\big(2d_h\!+\!\hat{t}_h^{k+1}\big)\!+\!\tau_h\!-\!\delta_h^1\tau_h\Big),\label{Eq.32a}\\
\underline{a}_{\ell}^k\!&=\!a_k\!+\!\sum_{h=k}^{\ell-1}\!\Big(\hat{b}_h^k\big(2d_h\!+\!\hat{t}_h^k\big)\!+\!\tau_h\!\!-\!\delta_h^1\tau_h\Big),\label{Eq.32b}
\end{align}
\end{subequations}
where $a_{k+1}$ and $a_k$ are the truck's arrival times at ramps $r_{k+1}$ and $r_k$, respectively. By \eqref{Eq.13}, $a_{k+1}$ is denoted as
\begin{align}
a_{k+1}=a_k\!+b_k^*(t_k^*\!+\!\hat{w}_k\!+\!2d_k)\!+\!\tau_k\!+\!\delta_{\tau,k},\label{Eq.33}
\end{align}
where $\hat{w}_k\geq{0}$ denotes the truck's actual waiting time at station $S_k$ and $\delta_{\tau,k}\!\in\![-\delta_k^1\tau_k,\delta_k^1\tau_k]$ denotes the actual value of the travel time uncertainty on $\tau_{k}$. Subsequently, by \eqref{Eq.32}-\eqref{Eq.33}, we obtain that
\begin{align}
\underline{a}_{\ell}^k-\underline{a}_{\ell}^{k+1}\!&=a_k\!-\!a_{k+1}\!+\!\sum_{h=k}^{\ell-1}\hat{b}_h^k\big(2d_h\!+\!\hat{t}_h^k\big)\!+\!\tau_k\!-\!\delta_k^1\tau_k\nonumber\\
&\quad-\!\!\sum_{h=k+1}^{\ell-1}\!\!\hat{b}_h^{k+1}\big(2d_h\!+\hat{t}_h^{k+1}\big)\nonumber\\
&\leq{a_k\!-a_{k+1}\!+b_k^*(2d_k\!+t_k^*)\!+\tau_k\!-\!\delta_k^1\tau_k}\nonumber\\
&=-b_k^*\hat{w}_k\!-\!(\delta_{\tau,k}\!+\delta_k^1\tau_k)\nonumber\\
&\leq{0}.\label{Eq.34}
\end{align}
This completes the proof of \eqref{Eq.20a}. 

Below, the case for $\hat{b}_{k+1}^{k+1}\!=\!0$ is discussed. Let us consider the more general case where 
\begin{align*}
\hat{b}_{k+1}^{k+1}=\hat{b}_{k+2}^{k+1}=\dots=\hat{b}_{i-1}^{k+1}=0,~\text{and}~\hat{b}_{i}^{k+1}\!=\!1.
\end{align*}
Similarly, we will show that 
\begin{align*}
\Big\{(b_k^*,t_k^*), & (\hat{b}_{k+1}^{k+1},\hat{t}_{k+1}^{k+1}), \dots,(\hat{b}_{i-1}^{k+1},\hat{t}_{i-1}^{k+1}),\nonumber\\
&(\hat{b}_{i}^{k+1},\hat{t}_{i}^{k+1}\!\!-\!\Delta{\hat{t}}_{i}),\big\{(\hat{b}_h^{k+1},\hat{t}_h^{k+1})\big\}_{h=i+1}^{\ell-1}\Big\}
\end{align*}
is feasible with respect to the constraints in \eqref{Eq.15} at ramp $r_k$, where $\Delta{\hat{t}}_{i}\!=\!(\delta_k^2\bar{P}\tau_k\!-\!\delta_{e,k})/{P_{i}^c}$ and $\hat{t}_i^{k+1}\!\geq\!{\Delta{\hat{t}_i}}$. By \eqref{Eq.29} and \eqref{Eq.31}, it can be seen that 
\begin{align*}
e_{h}^*-\hat{e}_{h}=\delta_k^2\bar{P}\tau_k-\delta_{e,k}\geq0,\quad h\!=\!k\!+\!1,\dots,i.
\end{align*}
Since $\hat{e}_{h}$, $h\!=\!k\!+\!1,\dots,i$, satisfy the constraint \eqref{Eq.15c}, $e_{h}^*$ also satisfy \eqref{Eq.15c}. The remaining parts can be proved using similar arguments as in the case where $\hat{b}_{k+1}^{k+1}\!=\!1$.
\item [(b)] Next, we will prove \eqref{Eq.20b}. To distinguish between \eqref{Eq.18} and \eqref{Eq.19} based on the truck's location, we denote by $\Delta\bar{e}_{k+1}^k$ the maximum increased battery at station $S_{k+1}$ computed at ramp $r_k$. By \eqref{Eq.17}, for $\ell\!=\!k\!+\!2,\dots,N$, we have that 
\begin{subequations}
\label{Eq.35}
\begin{align}
\!\!\!\bar{a}_{\ell}^{k+1}\!\!\!&=\!a_{k+1}\!\!+\!\!\Big(\!2d_{k+1}\!+\!\frac{\Delta\bar{e}_{k+1}^{k+1}}{P_{k+1}^c}\!+\!\bar{w}_{k+1}\!+\!\tau_{k+1}\!+\!\delta_{k+1}^1\tau_{k+1}\!\Big)\nonumber\\
&\quad+\!\!\sum_{h=k+2}^{\ell-1}\!\!\Big(2d_h\!+\!\frac{\Delta\bar{e}_h^{k+1}}{P_h^c}\!+\!\bar{w}_h\!+\!\tau_h\!+\!\delta_h^1\tau_h\Big),\label{Eq.35a}\\
\bar{a}_{\ell}^k\!\!&=\!a_k\!+\!\Big(2d_k\!+\!\frac{\Delta\bar{e}_k^k}{P_k^c}\!+\!\bar{w}_k\!+\!\tau_k\!+\!\delta_k^1\tau_k\Big)\nonumber\\
&\quad+\!\Big(2d_{k+1}\!+\!\frac{\Delta\bar{e}_{k+1}^k}{P_{k+1}^c}\!+\!\bar{w}_{k+1}\!+\!\tau_{k+1}\!+\!\delta_{k+1}^1\tau_{k+1}\Big)\nonumber\\
&\quad+\!\sum_{h=k+2}^{\ell-1}\!\!\Big(2d_h\!+\!\frac{\Delta\bar{e}_h^k}{P_h^c}\!+\!\bar{w}_h\!+\!\tau_h\!+\!\delta_h^1\tau_h\Big).\label{Eq.35b}
\end{align}
\end{subequations}
Note that $\Delta\bar{e}_h^k\!=\!\Delta\bar{e}_h^{k+1}$, for $h\!=\!k\!+\!2,\dots,\ell-1$, and $P_{k+1}^c\!=\!P_k^c$. Moreover, by \eqref{Eq.18} and \eqref{Eq.19}, $\Delta\bar{e}_k^k$, $\Delta\bar{e}_{k+1}^k$, and $\Delta\bar{e}_{k+1}^{k+1}$ are given by
\begin{align*}
    \Delta\bar{e}_k^k& = e_f\!-\!(\hat{e}_k\!-\!\bar{P}d_k),\\
    \Delta\bar{e}_{k+1}^k&=\bar{P}(d_{k}\!+\!\tau_{k}\!+\!d_{k+1})+\delta_{k}^2\bar{P}\tau_{k},\\
    \Delta\bar{e}_{k+1}^{k+1}&=e_f\!-\!(\hat{e}_{k+1}\!-\!\bar{P}d_{k+1}),
\end{align*}
where, by \eqref{Eq.8}, the actual battery energies $\hat{e}_k$ and $\hat{e}_{k+1}$ at ramps $r_k$ and $r_{k+1}$ satisfy
\begin{align*}
\hat{e}_k-\hat{e}_{k+1}=\bar{P}\tau_{k}+b^*_k(2\bar{P}d_k\!-\!t^*_kP_k^c)-\delta_{e,k}.
\end{align*}
It follows that
\begin{align*}
    &\Delta\bar{e}_{k+1}^{k+1}\!-\!\Delta\bar{e}_{k+1}^k\!-\!\Delta\bar{e}_k^k+b_k^*t_k^*P_k^c\\
    &=\hat{e}_k\!-\hat{e}_{k+1}\!-\!2\bar{P}d_k\!-\!\bar{P}\tau_{k}\!-\delta_{k}^2\bar{P}\tau_{k}+b_k^*t_k^*P_k^c\\
    &=-(\delta_{e,k}\!+\!\delta_{k}^2\bar{P}\tau_{k})+2\bar{P}d_k(b_k^*\!-\!1)\leq0.
\end{align*}
By \eqref{Eq.33}, subtracting \eqref{Eq.35b} from \eqref{Eq.35a} gives 
\begin{align}
\bar{a}_{\ell}^{k+1}\!\!-\!\bar{a}_{\ell}^k&=a_{k+1}\!-\!a_k+\frac{\Delta\bar{e}_{k+1}^{k+1}-\Delta\bar{e}_{k+1}^k}{P_{k+1}^c}\nonumber\\
&\quad-\Big(2d_k\!+\!\frac{\Delta\bar{e}_k^k}{P_k^c}\!+\!\bar{w}_k\!+\!\tau_k\!+\!\delta_k^1\tau_k\Big)\nonumber\\
&=b_k^*(t_k^*\!+\!\hat{w}_k\!+\!2d_k)\!+\!\delta_{\tau,k}\!-\!\big(2d_k\!+\!\bar{w}_k\!+\!\delta_k^1\tau_k\big)\nonumber\\
&\quad+\!\frac{\Delta\bar{e}_{k+1}^{k+1}\!-\!\Delta\bar{e}_{k+1}^k\!-\!\Delta\bar{e}_k^k}{P_k^c}\nonumber\\
&\leq{(\hat{w}_k\!-\!\bar{w}_k)+(\delta_{\tau,k}\!-\!\delta_k^1\tau_k)}\nonumber\\
&\quad+\frac{b_k^*t_k^*P_k^c+\Delta\bar{e}_{k+1}^{k+1}\!-\!\Delta\bar{e}_{k+1}^k\!-\!\Delta\bar{e}_k^k}{P_k^c}\nonumber\\
&\leq{0}.\label{Eq.36}
\end{align}
Note that, in \eqref{Eq.36}, $\hat{w}_k\!\leq\!{\bar{w}_k}$ and $\delta_{\tau,k}\!\in\![-\delta_k^1\tau_k,\delta_k^1\tau_k]$. Furthermore, $\Delta{\bar{e}_k^k}$ denotes the maximum increased battery of the truck at station $S_k$; cf. \eqref{Eq.18}. Therefore, $t_k^*P_k^c\!\leq\!{\Delta{\bar{e}_k^k}}$.  
\end{itemize}
With \eqref{Eq.34} and \eqref{Eq.36}, we conclude our proof.
\end{proofproposition1}

\begin{prooftheorem1}
Let $\{(b_{\ell}^{k^*}\!,t_{\ell}^{k^*})\}_{\ell=k}^N$ be the optimal solution of the problem \eqref{Eq.21}-\eqref{Eq.23} computed at ramp $r_k$. Recall that $\{\tilde{w}_{\ell}^k\}_{\ell=k}^N$ are the estimated waiting times provided by station $S_{\ell}$ to the truck when reaching ramp $r_k$ and $\{\hat{w}_{\ell}^k\}_{\ell=k}^N$ are the truck's actual waiting times at these stations.
\begin{itemize}
\item [(a)] By definition \eqref{Eq.21}, since $\max(a,b)\!=\!\frac{a+b+|a-b|}{2}$, we can rewrite $J_k^*$ and $\hat{J}_k$ for $k\!=\!1,\dots,N\!-\!1$ as 
\begin{subequations}
\label{Eq.37}
\begin{align}
\!\!\!J_k^*&=\xi{b_k^{k^*}}\!\big(2d_k\!+t_k^{k^*}\!\!+\!\tilde{w}_k^{k}\big)+\!\!\sum_{\ell=k+1}^N\!\!\xi{b_{\ell}^{k^*}}\!\big(2d_{\ell}\!+t_{\ell}^{k^*}\!\!+\!\tilde{w}_{\ell}^k\big)\nonumber\\
&\quad+\sum_{\ell=k}^N\theta_{\ell}b_{\ell}^{k^*}t_{\ell}^{k^*}\!+\!\frac{\gamma}{2}\big(a_{N+1}^{k^*}\!-\!a_{\text{end}}+|a_{N+1}^{k^*}\!-\!a_{\text{end}}|\big),\label{Eq.37a}\\
\!\!\!\hat{J}_k&=\xi{b_k^{k^*}}\!\big(2d_k\!+t_k^{k^*}\!\!+\!\hat{w}_k^k\big)+\!\!\sum_{\ell=k+1}^N\!\!\xi{b_{\ell}^{k^*}}\!\big(2d_{\ell}\!+t_{\ell}^{k^*}\!\!+\!\hat{w}_{\ell}^k\big)\nonumber\\
&\quad +\sum_{\ell=k}^N\theta_{\ell}b_{\ell}^{k^*}t_{\ell}^{k^*}\!+\!\frac{\gamma}{2}\big(\hat{a}_{N+1}^{k^*}\!-\!a_{\text{end}}+|\hat{a}_{N+1}^{k^*}\!-\!a_{\text{end}}|\big).\label{Eq.37b}
\end{align}
\end{subequations}
By \eqref{Eq.22e}, $a_{N+1}^{k^*}$ and $\hat{a}_{N+1}^{k^*}$ in \eqref{Eq.37} have the following form
\begin{align*}
a_{N+1}^{k^*}&=a_N^{k^*}\!+b_N^{k^*}\big(t_N^{k^*}\!\!+\!\tilde{w}_N^k\!+\!2d_N\big)+\tau_N\\
&=a_k+\sum_{\ell=k}^Nb_{\ell}^{k^*}\!\tilde{w}_{\ell}^k+\sum_{\ell=k}^N\big(b_{\ell}^{k^*}\!(t_{\ell}^{k^*}\!\!+2d_{\ell})+\tau_{\ell}\big),\\
\hat{a}_{N+1}^{k^*}&=\hat{a}_N^{k^*}\!+b_N^{k^*}\big(t_N^{k^*}\!\!+\!\hat{w}_N^k\!+\!2d_N\big)+\tau_N\\
&=a_k+\sum_{\ell=k}^Nb_{\ell}^{k^*}\!\hat{w}_{\ell}^k+\sum_{\ell=k}^N\big(b_{\ell}^{k^*}\!(t_{\ell}^{k^*}\!\!+\!2d_{\ell})+\tau_{\ell}\big).
\end{align*}
Since $\tilde{w}_k^k\!=\!\hat{w}_k^k$, by applying the inequalities $|a\!+\!b|\leq{|a|\!+\!|b|}$, $\big||a|\!-\!|b|\big|\leq{|a-b|}$ and based on the aforementioned derivations, we have 
\begin{align}
\big|\hat{J}_k\!-\!J_k^*\big|&\!=\!\Big|(\xi\!+\!\frac{\gamma}{2})\!\!\sum_{\ell=k+1}^N\!\! b_{\ell}^{k^*}\!(\hat{w}_{\ell}^k-\tilde{w}_{\ell}^k)\nonumber\\
&\quad~+\frac{\gamma}{2}\big(|\hat{a}_{N+1}^{k^*}\!-a_{\text{end}}|-|a_{N+1}^{k^*}\!-a_{\text{end}}|\big)\Big|\nonumber\\
&\leq{\!(\xi\!+\!\frac{\gamma}{2})\Big|\!\!\sum_{\ell=k+1}^N\!\!b_{\ell}^{k^*}\!\!(\hat{w}_{\ell}^k\!-\!\tilde{w}_{\ell}^k)\Big|}\!+\!\frac{\gamma}{2}\Big|\hat{a}_{N+1}^{k^*}\!-\!a_{N+1}^{k^*}\Big|\nonumber\\
&={(\xi\!+\!\gamma)\Big|\!\sum_{\ell=k+1}^N\!\!b_{\ell}^{k^*}\!\!(\hat{w}_{\ell}^k-\tilde{w}_{\ell}^k)\Big|}\nonumber\\
&\leq{(\xi\!+\!\gamma)\!\sum_{\ell=k+1}^N\!\!b_{\ell}^{k^*}\big|\hat{w}_{\ell}^k-\tilde{w}_{\ell}^k\big|}\nonumber\\
&\leq{(\xi\!+\!\gamma)\!\sum_{\ell=k+1}^N\!\!\Delta{w}_{\ell}^k},\label{Eq.38}
\end{align}
where $\Delta{w}_{\ell}^k\!=\!|\hat{w}_{\ell}^k\!-\!\tilde{w}_{\ell}^k|$ is defined as the error between the actual and estimated waiting times at station $S_{\ell}$. By \eqref{Eq.38}, we prove the upper bound of $|\hat{J}_k\!-\!J_k^*|$ for $k\!=\!1,\dots,N\!-\!1$. Particularly, as $\tilde{w}_N^N\!=\!\hat{w}_N^N$, there is $\hat{J}_N\!=\!J_N^*$. Thus, $\Delta{\bar{J}}_N\!=\!0$. 
\item [(b)] By \eqref{Eq.38}, the upper bound for $|\hat{J}_{k+1}\!-\!J_{k+1}^*|$ with $k\!=\!1,\dots,N\!-\!2$ is denoted as
\begin{align*}
\Delta{\bar{J}}_{k+1}=(\xi\!+\!\gamma)\!\!\sum_{\ell=k+2}^N\!\!\Delta{w}_{\ell}^{k+1}.
\end{align*}
In line with Proposition~\ref{Proposition 1}, we have $\Delta{w}_{\ell}^{k+1}\!\leq\!{\Delta{w}_{\ell}^k}$ for $\ell\!=\!k\!+\!2,\dots,N$. Thereby, we obtain that
\begin{align}
&\Delta{\bar{J}}_{k+1}-\Delta{\bar{J}}_k\nonumber\\
&=(\xi\!+\!\gamma)\Big(\sum_{\ell=k+2}^N\!\!\Delta{w}_{\ell}^{k+1}-\!\sum_{\ell=k+1}^N\!\!\Delta{w}_{\ell}^k\Big)\nonumber\\
&\leq{(\xi\!+\!\gamma)\Big(\sum_{\ell=k+2}^N\!\!\Delta{w}_{\ell}^k-\!\sum_{\ell=k+2}^N\!\!\Delta{w}_{\ell}^k-\Delta{w}_{k+1}^k\Big)}\nonumber\\
&\leq{0}.\label{Eq.39}
\end{align} 
For $k\!=\!N\!-\!1$, since $\Delta{\bar{J}}_N=0$ and $\Delta{w}_{\ell}^k\geq{0}$, we have $\Delta{\bar{J}}_N-\Delta{\bar{J}}_{N-1}=-(\xi\!+\!\gamma)\Delta{w}_N^{N-1}\!\leq\!{0}$. This completes the proof of Theorem~\ref{Theorem 1}.
\end{itemize}
\end{prooftheorem1}

\bibliographystyle{IEEEtran}
\bibliography{Ref}

\begin{IEEEbiography}[{\includegraphics[width=1in,height=1.25in,clip,keepaspectratio]{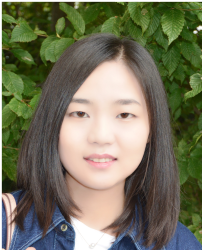}}]{\textcolor{black}{Ting Bai}}\textcolor{black}{received the B.Sc. degree in automation from Northwestern Polytechnical University, Xi'an, China, in 2013, and the Ph.D. degree in electrical engineering from Shanghai Jiao Tong University, Shanghai, China, in 2019. From 2020 to 20204, she was a Postdoctoral Researcher at the Department of Electrical Engineering and Computer Science, KTH Royal Institute of Technology, Stockholm, Sweden. Since 2024, she has been a Postdoctoral Research Associate with the School of Civil and Environmental Engineering at Cornell University, Ithaca, New York, USA. Her research interests include distributed model predictive control, dynamic programming, reinforcement learning, and optimal control of large transportation systems.} 
\end{IEEEbiography}
\vspace{-20pt}

\begin{IEEEbiography}[{\includegraphics[width=1in,height=1.25in,clip,keepaspectratio]{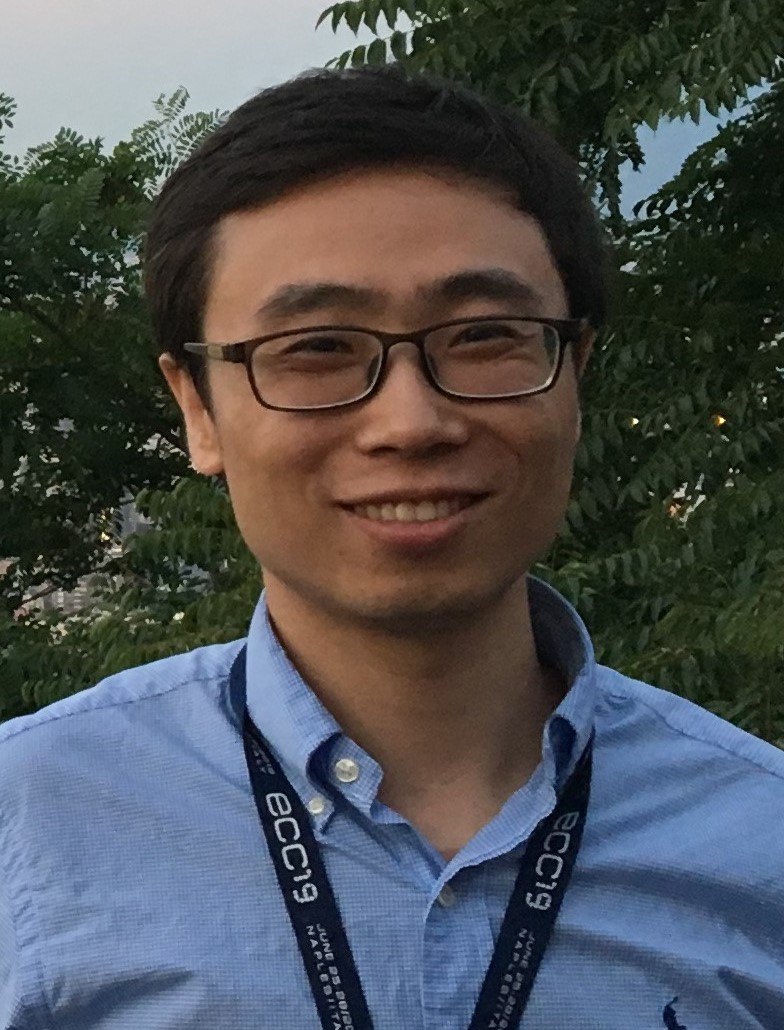}}]{Yuchao Li} received the B.Eng. degree in mechanical engineering from Harbin Institute of Technology, Harbin, China, in 2015. He received the M.Sc. degree in mechatronics and the Ph.D. degree in electrical engineering both from the Royal Institute of Technology, Stockholm, Sweden, in 2017 and 2023, respectively. Since 2023, he has been a Postdoctoral Research Scholar with the Arizona State University, Tempe, USA. His research interests include optimal control, reinforcement learning, network optimization, and their applications to transportation systems. 
\end{IEEEbiography}
\vspace{-20pt}

\begin{IEEEbiography}[{\includegraphics[width=1in,height=1.25in,clip,keepaspectratio]{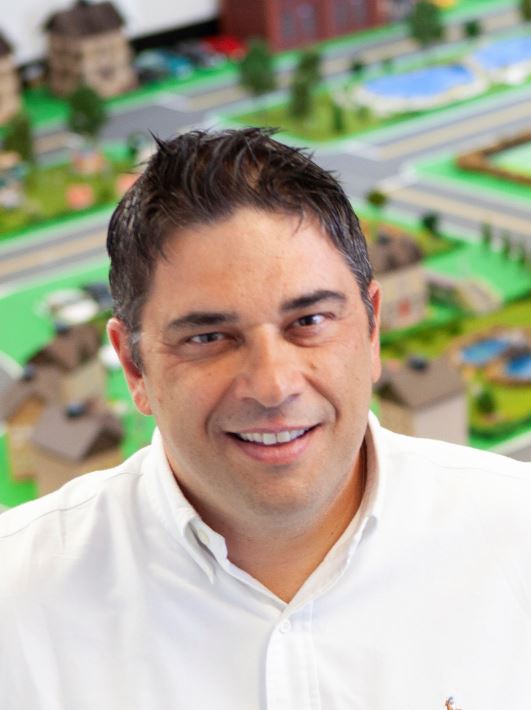}}]{Andreas A. Malikopoulos} (S'06--M'09--SM'17) received a Diploma in mechanical engineering from the National Technical University of Athens (NTUA), Greece, in 2000. He received M.S. and Ph.D. degrees in mechanical engineering at the University of Michigan, Ann Arbor, Michigan, USA, in 2004 and 2008, respectively. He is a professor in the School of Civil and Environmental Engineering at Cornell University and the director of the Information and Decision Science (IDS) Laboratory.
Prior to these appointments, he was the Terri Connor Kelly and John Kelly Career Development Professor in the Department of Mechanical Engineering (2017-2023) and the founding Director of the Sociotechnical Systems Center (2019-2023) at the University of Delaware (UD). Before he joined UD, he was the Alvin M. Weinberg Fellow (2010-2017) in the Energy \& Transportation Science Division at Oak Ridge National Laboratory (ORNL), the Deputy Director of the Urban Dynamics Institute (2014-2017) at ORNL, and a Senior Researcher in General Motors Global Research \& Development (2008-2010). His research spans several fields, including analysis, optimization, and control of cyber-physical systems (CPS); decentralized stochastic systems; stochastic scheduling and resource allocation; and learning in complex systems. His research aims to develop theories and data-driven system approaches at the intersection of learning and control for making CPS able to realize their optimal operation while interacting with their environment. He has been an Associate Editor of the IEEE Transactions on Intelligent Vehicles and IEEE Transactions on Intelligent Transportation Systems from 2017 through 2020. He is currently an Associate Editor of Automatica and IEEE Transactions on Automatic Control, and a Senior Editor of IEEE Transactions on Intelligent Transportation Systems. He is a member of SIAM, AAAS, and a Fellow of the ASME.
\end{IEEEbiography}
\vspace{-20pt}

\begin{IEEEbiography}[{\includegraphics[width=1in,height=1.25in,clip,keepaspectratio]{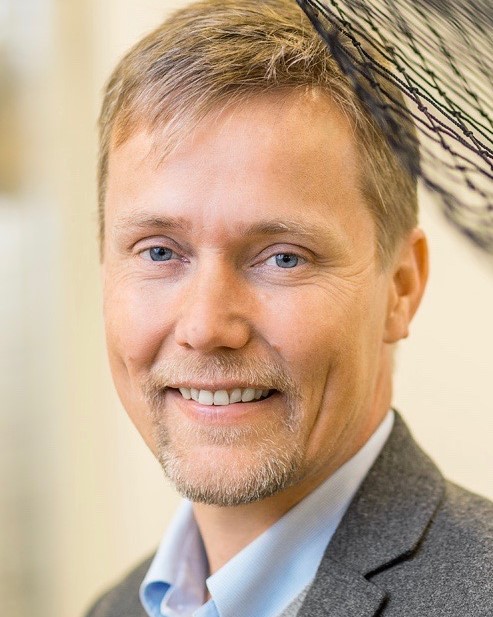}}]{Karl Henrik Johansson} is Swedish Research Council Distinguished Professor in Electrical Engineering and Computer Science at KTH Royal Institute of Technology in Sweden and Founding Director of Digital Futures. He earned his M.Sc. degree in Electrical Engineering and Ph.D. in Automatic Control from Lund University. He has held visiting positions at UC Berkeley, Caltech, NTU, and other prestigious institutions. His research interests focus on networked control systems and cyber-physical systems with applications in transportation, energy, and automation networks. For his scientific contributions, he has received numerous best paper awards and various distinctions from IEEE, IFAC, and other organizations. He has been awarded Distinguished Professor by the Swedish Research Council, Wallenberg Scholar by the Knut and Alice Wallenberg Foundation, Future Research Leader by the Swedish Foundation for Strategic Research. He has also received the triennial IFAC Young Author Prize and IEEE CSS Distinguished Lecturer. He is the recipient of the 2024 IEEE CSS Hendrik W. Bode Lecture Prize. His extensive service to the academic community includes being President of the European Control Association, IEEE CSS Vice President Diversity, Outreach $\&$ Development, and Member of IEEE CSS Board of Governors and IFAC Council. He has served on the editorial boards of Automatica, IEEE TAC, IEEE TCNS and many other journals. He has also been a member of the Swedish Scientific Council for Natural Sciences and Engineering Sciences. He is Fellow of both the IEEE and the Royal Swedish Academy of Engineering Sciences. 
\end{IEEEbiography}
\vspace{-20pt}

\begin{IEEEbiography}[{\includegraphics[width=1in,height=1.25in,clip,keepaspectratio]{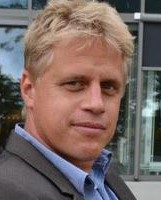}}]{Jonas M{\aa}rtensson} received the M.Sc. degree in vehicle engineering and the Ph.D. degree in automatic control from KTH Royal Institute of Technology, Stockholm, Sweden, in 2002 and 2007, respectively. In 2016, he was appointed as a Docent. He is currently a Professor with the Division of Decision and Control Systems, at KTH Royal Institute of Technology. He is also engaged as the Director of the Integrated Transport Research Laboratory (ITRL) and the Thematic Leader for the area of transport in the information age with the KTH Transport Platform. His research interests span efficient and sustainable transport systems, including cooperative traffic control and control of connected and automated vehicles, with a particular interest in freight and heavy vehicles.
\end{IEEEbiography}
\end{document}